\documentclass[twocolumn]{aastex62}

\hypersetup{linkcolor=red}
\graphicspath{{./}{Figures/}}

\newcommand\st{\bgroup\markoverwith
{\textcolor{magenta}{\rule[0.5ex]{2pt}{1pt}}}\ULon}

\usepackage{amsmath, bm, hyperref, ulem}
\usepackage{xcolor, colortbl, array, makecell} 
\newcolumntype{?}[1]{!{\vrule width #1}}
\usepackage{savesym} \savesymbol{tablenum}
\usepackage[detect-weight=true, detect-family=true]{siunitx}
\usepackage{hyperref}
\hypersetup{
    colorlinks=true,
    linkcolor=cyan,
    filecolor=magenta,      
    urlcolor=blue,
    pdftitle={Overleaf Example},
    pdfpagemode=FullScreen,
    }
\restoresymbol{SIX}{tablenum}
\usepackage[T1]{fontenc}
\usepackage[overload]{textcase}
\usepackage{xcolor}
\usepackage{totcount}
\usepackage{lineno}

\newtotcounter{citnum} 
\def\oldbibitem{} \let\oldbibitem=\bibitem
\def\bibitem{\stepcounter{citnum}\oldbibitem}
\graphicspath{{./}{Figures/}}
\shorttitle{WASP-25 and WASP-125 Transmission Spectra} \shortauthors{McGruder et al.}

\begin{document}

\title{ACCESS, LRG-BEASTS, \& MOPSS: Featureless Optical Transmission Spectra of WASP-25b and WASP-124b}

\correspondingauthor{Chima D. McGruder} \email{chima.mcgruder@cfa.harvard.edu}

\author[0000-0002-6167-3159]{Chima D. McGruder}\affiliation{Center for Astrophysics ${\rm \mid}$ Harvard {\rm \&} Smithsonian, 60 Garden St, Cambridge, MA 02138, USA}

\author[0000-0003-3204-8183]{Mercedes L\'opez-Morales} \affiliation{Center for Astrophysics ${\rm \mid}$ Harvard {\rm \&} Smithsonian, 60 Garden St, Cambridge, MA 02138, USA}

\author[0000-0002-4207-6615]{James Kirk}\affiliation{Department of Physics, Imperial College London, Prince Consort Road, London, SW7 2AZ, UK}

\author[0000-0002-3627-1676]{Benjamin V.\ Rackham} \altaffiliation{51 Pegasi b Fellow}
\affiliation{Department of Earth, Atmospheric and Planetary Sciences, Massachusetts Institute of Technology, 77 Massachusetts Avenue, Cambridge, MA 02139, USA}
\affiliation{Kavli Institute for Astrophysics and Space Research, Massachusetts Institute of Technology, Cambridge, MA 02139, USA}

\author[0000-0002-2739-1465]{Erin May}\affiliation{Bloomberg Center for Physics and Astronomy, 3400 N. Charles Street, Baltimore, MD 21218, USA}

\author[0000-0003-0973-8426]{Eva-Maria Ahrer} \affiliation{Centre for Exoplanets and Habitability, University of Warwick, Gibbet Hill Road, CV4 7AL Coventry, UK}
\affiliation{Department of Physics, University of Warwick, Gibbet Hill Road, CV4 7AL Coventry, UK}

\author[0000-0002-3641-6636]{George W. King} \affiliation{Department of Astronomy, University of Michigan, Ann Arbor, MI 48109, USA}
\affiliation{Department of Physics, University of Warwick, Gibbet Hill Road, CV4 7AL Coventry, UK}
\affiliation{Centre for Exoplanets and Habitability, University of Warwick, Gibbet Hill Road, CV4 7AL Coventry, UK}

\author[0000-0003-4157-832X]{Munazza K. Alam}
\affiliation{Carnegie Earth {\rm \&} Planets Laboratory, 5241 Broad Branch Road NW, Washington, DC 20015, USA}

\author[0000-0002-0832-710X]{Natalie H. Allen} \affiliation{William H. Miller III Department of Physics and Astronomy, Johns Hopkins University, 3400 N Charles St, Baltimore, MD 21218, USA}

\author[0000-0003-3455-8814]{Kevin Ortiz Ceballos}\altaffiliation{NSF Graduate Research Fellow}\affiliation{Center for Astrophysics ${\rm \mid}$ Harvard {\rm \&} Smithsonian, 60 Garden St, Cambridge, MA 02138, USA}

\author[0000-0001-9513-1449]{N\'estor Espinoza} \affiliation{Space Telescope Science Institute (STScI), 3700 San Martin Dr, Baltimore, MD 21218, USA}

\author[0000-0002-3003-3183]{Tyler Gardner} \affiliation{Astrophysics Group, Department of Physics \& Astronomy, University of Exeter, Stocker Road, Exeter, EX4 4QL, UK}

\author[0000-0002-5389-3944]{Andr\'es Jord\'an}
\affiliation{Facultad de Ingeniera y Ciencias, Universidad Adolfo Ib\'{a}\~{n}ez, Av. Diagonal las Torres 2640, Pe\~{n}alol\'{e}n, Santiago, Chile}
\affiliation{Millennium Institute for Astrophysics, Chile} 
\affiliation{Data Observatory Foundation, Chile}

\author{Kelly Meyer} \affiliation{University of Michigan, Astronomy Department, Ann Arbor, MI 48109 USA}

\author[0000-0002-3380-3307]{John D. Monnier} \affiliation{University of Michigan, Astronomy Department, Ann Arbor, MI 48109 USA}

\author[0000-0003-0412-9664]{David J. Osip}\affiliation{Las Campanas Observatory, Carnegie Institution of Washington, Colina el Pino, Casilla 601 La Serena, Chile}

\author[0000-0003-1452-2240]{Peter J. Wheatley} \affiliation{Centre for Exoplanets and Habitability, University of Warwick, Gibbet Hill Road, CV4 7AL Coventry, UK}
\affiliation{Department of Physics, University of Warwick, Gibbet Hill Road, CV4 7AL Coventry, UK}




\turnoffedit To turn off the edits, from https://journals.aas.org/revision-history/
\begin{abstract} 
We present new optical transmission spectra for two hot Jupiters: WASP-25b (M~=~0.56~M$_J$; R~=~1.23~R$_J$; P~=~3.76~days) and WASP-124b (M~=~0.58~M$_J$; R~=~1.34~R$_J$; P~=~3.37~days), with wavelength coverages of 4200 -- 9100\AA~and 4570 -- 9940\AA, respectively. These spectra are from the ESO Faint Object Spectrograph and Camera (v.2) mounted on the New Technology Telescope (NTT) and Inamori-Magellan Areal Camera \& Spectrograph on Magellan Baade. No strong spectral features were found in either spectra, with the data probing 4 and 6 scale heights, respectively. \texttt{Exoretrievals} and \texttt{PLATON} retrievals favor stellar activity for WASP-25b, \edit3{while the data for WASP-124b did not favor one model over another}. For both planets the retrievals found a wide range in the depths where the atmosphere could be optically thick \edit2{($\sim$0.4~$\mu$ -- 0.2~bars for WASP-25b and 1.6~$\mu$ -- 32~bars for WASP-124b)} and recovered a temperature that is consistent with the planets' equilibrium temperatures, but with wide uncertainties \edit2{(up to $\pm$430K)}. For WASP-25b, the models also favor stellar spots that are \edit3{$\sim$500--3000K} cooler than the surrounding photosphere. The fairly weak constraints on parameters are owing to the relatively low precision of the data, with an average precision of 840 and \edit3{1240}~ppm per bin for WASP-25b and WASP-124b, respectively. 
However, some contribution might still be due to an inherent absence of absorption or scattering in the planets’ upper atmospheres, possibly because of aerosols. We attempt to fit the strength of the sodium signals to the aerosol--metallicity trend proposed by \cite{McGruder:2023}, and find WASP-25b and WASP-124b are consistent with the prediction, though their uncertainties are too large to confidently confirm the trend.
\end{abstract}

\keywords{planets and satellites: atmospheres --- 
stars: activity; starspots --- techniques: spectroscopic; WASP-25b; WASP-124b}


\section{Introduction} \label{sec:intro4}
Exoplanet science is at a new frontier, where the need to repurpose telescopes to observe planetary atmospheres is being replaced with telescopes that are designed with the goal of exoplanet atmosphere characterization in mind. Exoplanet scientists have made plenty of advancements with the current generation of telescopes. Specifically with low-resolution transmission spectroscopy, strives have been made with ground-based telescopes \citep[e.g.][]{Sing2012, Nikolov2016, Diamond-Lowe2018, Todorov2019, Spyratos2021}, the Hubble Space Telescope \citep[HST; e.g.][]{Charbonneau:2002, Kulow2014, Tsiaras2016, Sing:2016, Wakeford2020, Rathcke2021}, and Spitzer \citep[e.g.][]{Knutson2011, Pont2013, Alam2020, Alderson2022}. However, the advancements with the next generation of telescopes will be revolutionary. This has already begun with the launch and utilization of the JWST, where novel science has been conducted with outstanding quality of data and newly discovered molecular features \citep[e.g.][]{Tsai2022, Ahrer2023, Feinstein2023, Rustamkulov2023, Alderson2023}. Furthermore, soon-to-be-launched telescopes like Pandora \citep{Quintana:2021, Hoffman:2022}, 
the Atmospheric Remote-sensing Infrared Exoplanet Large-survey \citep[ARIEL; ][]{Tinetti2018}, 
and the next generation of ground-based telescopes: the Extremely Large Telescope (ELT) \footnote{ELT:\href{https://elt.eso.org/}{https://elt.eso.org/}}, 
Thirty Meter Telescope (TMT) \footnote{TMT:\href{https://www.tmt.org/}{https://www.tmt.org/}}, 
and Giant Magellan Telescope (GMT) \footnote{GMT:\href{https://giantmagellan.org/}{https://giantmagellan.org/}} 
will have designs and instruments specific for exoplanet atmospheric studies. These telescopes will undoubtedly be cornerstones in advancing our understanding of exoplanet atmospheres. 

There is still much about exoplanet atmospheres that alludes us. For example, we have no direct link with physical conditions that cause high-altitude aerosols to form in the upper atmosphere of some observed planets \citep[e.g.][]{Alam2018, Chachan2019, Estrela2021} but not others \citep[e.g.][]{Sing:2016,Kirk:2019,Alam:2021,Ahrer:2022,McGruder:2022}. Here we refer to aerosols as clouds--- condensation material due to specific atmospheric conditions---or hazes---material formed from photochemical reactions. The formation of aerosols likely occurs in most (if not all) atmospheres, just like in our solar system; however, high-altitude aerosols are normally the main concern when probing atmospheres. This is because the most used method to probe exoplanet atmospheres is transmission spectroscopy, which more easily probes the upper atmospheric limbs of planets, due to geometry and opacities \citep{Lecavelier2008, Sing:2018, Kreidberg2018}. There are a number of studies aimed at understand the composition and formation of high-altitude hazes \citep[e.g.][]{moses:2011,moses:2013,Fleury:2019} and clouds \citep[e.g.][]{Helling:2019, Gao:2020,Estrela:2022}, 
and though there has been a lot of headway toward this, there is little observational support for leading theories. 

Additionally, finding observational trends in aerosol formation has proven illusive, with many contradicting or inconclusive studies \citep{Heng:2016,Stevenson:2016,Fu:2017,Tsiaras:2018,Fisher:2018,Dymont2021,Estrela:2022}. A possible reason why no correlation has been clearly identified is the limited number of observed planets relative to the parameter space, where tens of planetary atmospheres have been used for studies, but tens of parameters (host star, orbital, and planetary parameters) could be correlated to aerosol formation rates. Possible solutions are to either increase the number of planetary atmospheres observed or to reduce the parameter space by observing select targets with many parameters that nearly match. The observations presented here aim to address both methods. 

We observed the atmospheres of WASP-25b \citep[M = 0.6 M$_J$, R = 1.2 R$_J$, P = 3.764 d, host star = G4, V$_{mag}$ = 11.9][]{Enoch2011, Brown2012, Southworth2014}, and WASP-124b \citep[M = 0.6 M$_J$, R = 1.3 R$_J$, P = 3.373 d, host star = F9, V$_{mag}$ = 12.7]{Maxted2016}. We obtained three spectroscopic transits of WASP-25b and five of WASP-124b as part of ACCESS.\footnote{The Atmospheric Characterization Collaboration for Exoplanet Spectroscopic Studies survey on the Baade Magellan Telescope \citep{Jordan:2013, Rackham:2017, Bixel:2019, Espinoza2019, Weaver:2020, McGruder:2020, Weaver:2021, Kirk:2021, McGruder:2022, Allen2022}}. We obtained one additional transit of WASP-25b with the ESO Faint Object Spectrograph and Camera (v.2; EFOSC2) instrument on the ESO New Technology Telescope (NTT) as part of 
LRG-BEASTS\footnote{The Low Resolution Ground-Based Exoplanet Atmosphere Survey using Transmission Spectroscopy \citep{Kirk:2017, Kirk:2018, Kirk:2019,Kirk:2021,Louden:2017,Alderson:2020,Ahrer:2022,Ahrer:2023b}}. There was also a full and partial transit of WASP-124b obtained by the MOPSS team\footnote{Michigan/Magellan Optical Planetary Spectra Survey \citep{May:2018, May:2020}} that we add to our dataset. Neither of these planets have atmospheric observations published. Furthermore, they have very similar parameters to one another and are part of a sample of seven planets proposed by \cite{McGruder:2023} that could be systems key for identifying correlations with high-altitude aerosols and observed parameters. 

The observation and reduction of all data are described in Section \ref{sec:Observations}, followed by their light curve analysis in Section \ref{sec:LC_analysis}. In Section \ref{sec:TransSpec} we introduce the combined transmission spectra of both targets, discuss our retrieval analysis of the data (Section \ref{sec:Retrievals}), and interpret the retrieval results (Section \ref{sec:RetrivInterp}). We then compare our results with expectations from the tentative aerosol--metallicity trend proposed by \cite{McGruder:2023} in Section \ref{sec:Sim7}. Finally, in Section \ref{sec:Sum+Conc} we recapitulate and provide conclusions. 

\section{Observations and Data Reduction} \label{sec:Observations}
\subsection{Magellan/IMACS Transits} \label{Magellan_Observations}
We observed three transits of WASP-25b (UTYYMMDD: UT180620, UT210306, UT220325) and five transits of WASP-124b (UT190826, UT210809, UT210905, UT211002, UT220605) with the Inamori-Magellan Areal Camera \& Spectrograph \cite[IMACS;][]{2011Dressler} mounted on Baade, one of the twin 6.5-m Magellan telescopes. Those transits were observed as part of ACCESS and used a setup similar to previous ACCESS observations \citep[i.e.][]{Weaver:2021,Kirk:2021,McGruder:2022,Allen2022}. Our general setup uses the 8K\,$\times$\,8K CCD mosaic camera at the f/2 focus, a 300\,line/mm grating at blaze angle of 17.5$^{\circ}$, and a GG455 filter. This gave a wavelength coverage of (4550--9900\,\AA) and dispersed the spectra across two chips, but we managed to fit the spectrum of the target from 4550--9100\,\AA{} on one CCD, preventing gaps from occurring at wavelengths of particular interest (see Figure \ref{fig:ExtraSpec}). We used 2\,$\times$\,2 binning and the FAST readout mode to reduce readout time and improve observational duty cycle. We used $\SI{10}{\arcsecond}$ by $\SI{90}{\arcsecond}$ slits ($\SI{0.5}{\arcsecond}$ by $\SI{90}{\arcsecond}$ for HeNeAr wavelength calibration lamps), putting the observations in a seeing-limited regime, with an average resolving power of R = 1350. The number of exposures, range of airmasses, \edit2{instrument setup}, and resolution per night can be found in Table \ref{tab:ObsLog+CompStars}.

We combined our ACCESS observations of WASP-124b with two observations obtained by the MOPSS team on UT180915 and UT190615 (UT190615 was a partial transit). We used Magellan/IMACS with a similar observational set up to ACCESS's, but with the 300 line/mm grating at a blaze angle of 26.7$^{\circ}$ and a slit size of $\SI{15}{\arcsecond}$ by $\SI{20}{\arcsecond}$ \edit3{($\SI{1}{\arcsecond}$ by $\SI{1}{\arcsecond}$ for HeNeAr wavelength calibration lamps)}. 
We also had different comparison stars, which are highlighted in Table \ref{tab:ObsLog+CompStars}.

All IMACS observations utilize the multiobject spectrograph (MOS) mode to observe multiple comparison stars simultaneously. The best comparison stars were selected based on the procedure outlined in \cite{Rackham:2017}, where we consider a nearby star suitable if it had a color difference of $D < 1$ with the target. $D$ is defined as
\begin{equation*}
D=\sqrt{[(B-V)_c - (B-V)_t]^2 + [(J-K)_c - (J-K)_t]^2},
\end{equation*}
where the uppercase letters correspond to the Johnson-Cousin apparent magnitudes of the stars, and the subscripts $t$ and $c$ indicate the target and potential comparison, respectively. The sky coordinates of each comparison, and their $D$ relative to the target, can be found in Table \ref{tab:ObsLog+CompStars}.

\subsection{IMACS Reduction} \label{IMACS_Reduction}
The reduction process for all IMACS data was the same, and uses a custom ACCESS pipeline which has been described in detail by \cite{Jordan:2013, Espinoza:2017}, \cite{Rackham:2017} and, \cite{Bixel:2019}. In general, this includes wavelength calibration using the HeNeAr arc lamp measurements, bias subtraction with the overscan region, pixel tracing, and sky background subtraction utilizing the median counts outside the science aperture. \edit2{The radius of the science aperture was determined by taking the average full width half max (FWHM) over time and wavelength. This value was then added to three times the standard deviation (STD) of all FWHM values (all calculated wavelength and time dependent FWHM values), i.e. aperture = <FWHM> + 3$\times$STD$_{FWHM}$.} \cite{Allen2022} found that optimal extraction \citep{Marsh1989OptExtract} has the potential to be a more effective way to identify and correct for bad-pixels and cosmic-rays in ACCESS data. When testing the effectiveness of this reduction step versus the traditional pipeline steps, we see slightly less scatter in the resulting white light curve with optimal extraction, so we adopt the results of this method. The final reduced spectra for the targets from each night is shown in Figure \ref{fig:ExtraSpec}.

\subsection{NTT/EFOSC2 Transit} \label{NTT_Observations}
WASP-25b had an additional transit observation with the European Faint Object Spectrograph \& Camera 2 \citep[EFOSC2; ][]{Buzzoni:1984} mounted on the 3.6-m New Technology Telescope (NTT) as part of LRG-BEASTS and ESO program 0100.C-0822(A) (PI: Kirk). The observation was taken on the night of UT180329 with the same instrument and setup used to detect Na in the atmosphere of WASP-94Ab \citep{Ahrer:2022} and clouds in the atmosphere of HATS-46b \citep{Ahrer:2023b}. This was a 27\,\arcsec $\times$ 3.53\,\arcmin~slit and Grism \#11, which provided spectra from 3960--7150\,\AA~and an average seeing-limited resolution of R = 150, which was dispersed on a 2048\,$\times$\,2048 pixel Loral/Lesser CCD. Our long slit allowed us to simultaneously observe the comparison star 2MASS J13011275-2730485 with a $V = 12.5$ and $B-V=1.5$, where WASP-25 has a $V = 11.9$ and $B-V=0.7$. We used $2\times2$ binning and the fast readout mode.

We also obtained 54 biases, 7 lamp flats, 17 morning twilight sky flats\footnote{A communication problem with the instrument prevented us from obtaining additional lamp flats.} and 3 arc lamps at the beginning and end of the observations. The flats were taken with the same 27\,\arcsec-wide slit as the science observations but the arc lamps were taken with a 1\,\arcsec~slit to avoid saturation and ensure narrow lines for calibration. This meant that the arc lamps were only used for an initial wavelength calibration with the final wavelength calibration performed using absorption lines in the stellar spectra. 

\subsection{EFOSC2 Reduction} \label{EFOSC_Reduction}
We reduced and processed the data using the LRG-BEASTS pipeline which is described in more detail in \cite{Kirk:2017,Kirk:2018,Kirk:2021}. For the flats, we created two sets of reductions, one without a flat-field correction, as is standard for LRG-BEASTS observations \citep[e.g.,][]{Alderson:2020,Kirk:2021,Ahrer:2022}, and one with a flat-field correction. This flat-field correction was performed in a novel way whereby we used the master sky flat without removing the sky spectrum from the sky flat. The motivation behind this approach was to avoid uncertainties associated with fitting out the sky spectrum (with a running median for example) while also capitalizing on the higher number of blue photons from the sky flat compared to a lamp flat. Since we did not remove the sky spectrum from the sky flat, this meant that the stellar spectra ($F_1$, $F_2$) we extracted were contaminated by the sky background imprinted into the sky flat ($F_\mathrm{sky}$). \edit2{Therefore, the stellar spectra we extracted were actually $F_1/F_\mathrm{sky\textcolor{red}{,1}}$ and $F_2/F_\mathrm{sky,\textcolor{red}{2}}$. The target and reference stars drifted across the course of the observations, leading to changes in $F_\mathrm{sky,1}$ and $F_\mathrm{sky,2}$. This meant that the $F_\mathrm{sky}$ terms did not fully cancel out after dividing the target's light curve ($\Sigma(F_1/F_\mathrm{sky,1})$) by the comparison's light curve ($\Sigma(F_2/F_\mathrm{sky,2})$). However, we found that using the sky flat led to light curves and transmission spectra that deviated by $<< 1\sigma$ from the same reduction without the flat field, while also decreasing the white noise in the light curves by 6\,\% and leading to a small improvement the precision in the transmission spectrum ($\sim$3\,\%). This demonstrates that impacts on the spectrophotometry due to $\Delta(F_\mathrm{sky,1}$) and $\Delta(F_\mathrm{sky,2}$) are insignificant, likely due to the fact that the stellar traces drift by $< 3$ pixels, which is corrected for by cross-correlation, and this constitutes only 8\,\% of our average bin width used to make the transmission spectrum. Due to this test, we adopted the reduction using the sky flat for the rest of our analysis.}

\begin{figure*}[htb]
    \centering
    \includegraphics[width=1\textwidth]{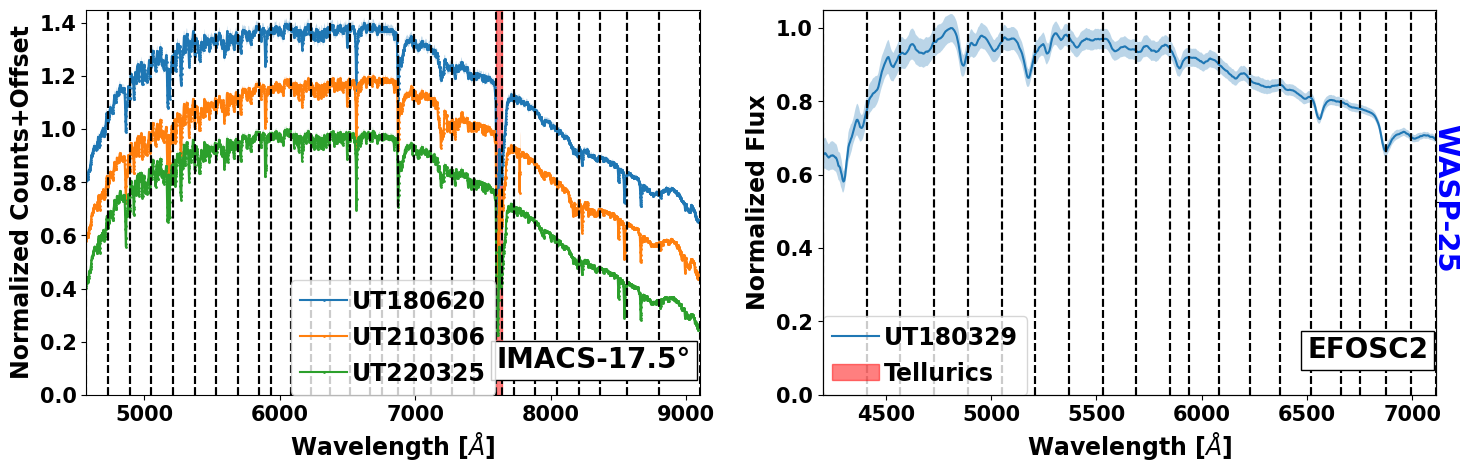}
    \includegraphics[width=1\textwidth]{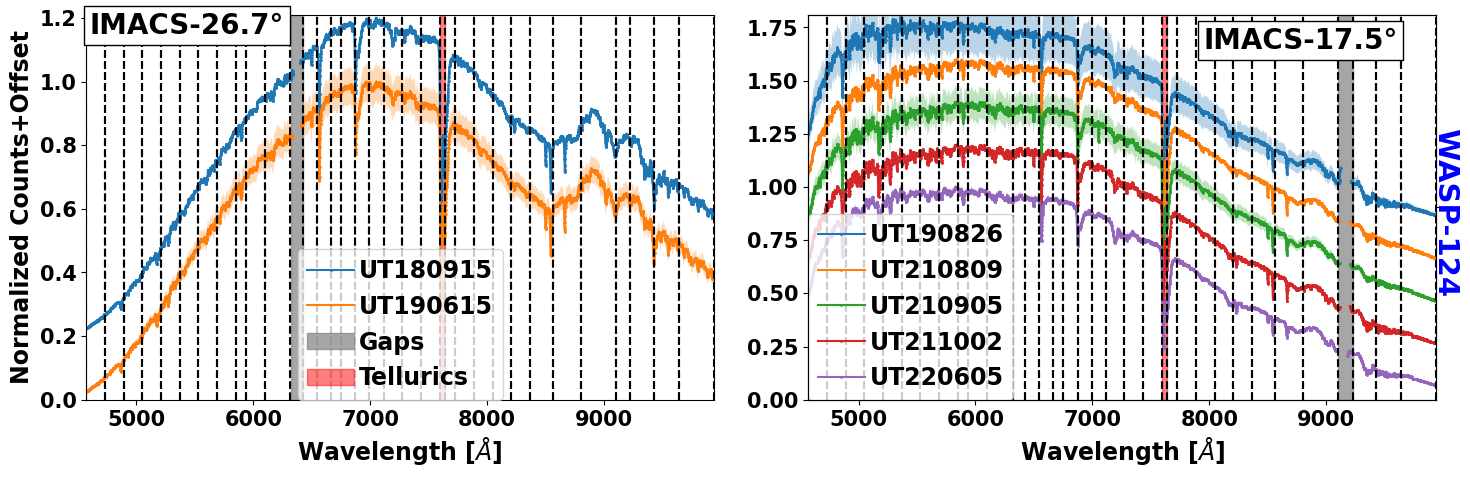}
    \caption{Median extracted spectra of WASP-25 (top) and WASP-124 (bottom). Each spectra is plotted in the same order in which it is printed in the legend (top to bottom). The shaded regions of the same color extending past the median lines are the 1$\sigma$ range of counts extracted for that night. Each spectroscopic bin used is demarcated by dotted vertical lines, where the only gaps in the binning scheme are the strong telluric region from 7594--7638\AA{} (lightly red shaded region), the CCD gap for the first two WASP-124 observations (bottom left) at 6317--6424\AA, and the CCD gap for the other five WASP-124 observations (bottom right) at 9100--9225\AA (gray shaded region). The specific instrument and setup used for each spectra is printed, where EFOSC2, IMACS-17.5$^{\circ}$, and IMACS-26.7$^{\circ}$ refer to the LRG-BEASTS (see Section \ref{NTT_Observations}), ACCESS, and MOPSS (see Section \ref{Magellan_Observations}) setups, respectively. The differing throughputs between the grism used for the ACCESS and MOPSS data explains the different spectral shapes between the two WASP-124 spectra on the bottom left and the five spectra on the bottom right. Note the plotted spectrum of the NTT/EFOSC2 data (top right), was created without the sky flat, in order to visually compare with the other spectra. However, the final spectrum used for data analysis did use the sky flat as discussed in Section \ref{EFOSC_Reduction}. } 
    \label{fig:ExtraSpec} 
\end{figure*}

To extract the stellar spectra we experimented with different aperture widths and background widths and compared the noise of the resulting white light curve in each case. We found the lowest white light noise resulted from a combination of a 22-pixel-wide aperture, with two 15-pixel-wide background regions on either side of the aperture, each separated from the aperture by 15 pixels. We fit a linear polynomial across these two background regions to remove the sky background from our spectra. Following the extraction of the stellar spectra, we clipped cosmic ray hits via identifying $5\sigma$ outliers in the spectral time series and replaced these with a linear interpolation between the nearest two neighboring pixels. We then corrected for shifts in the stellar spectra of $\pm 2$ pixels across the night. This was done by cross-correlating each spectrum with a spectrum taken in the middle of the observations and then performing a flux-conserving resampling of each spectrum onto the reference wavelength grid. Figure \ref{fig:ExtraSpec} (top right) shows the final spectrum extracted for WASP-25 with this pipeline.

\newcommand\ChangeRT[1]{\noalign{\hrule height #1}}
\begin{deluxetable*}{cccccccc}[htb]
    \caption{Observing log for the WASP-25b and WASP-124b data sets}
    \label{tab:ObsLog+CompStars}
    \tablehead{{Transit Date}  & {Instrument} & {Airmass} & {Exposure} & {Frames} &{resolution}  & {Comparisons' Coordinates} & {Color Difference,}\\ {(UTC)} & {set-up} & {[range]}  & {Times (s)} & &{[min/max]} & {[RA, Dec]} &{($D$)}} 
    \startdata 
     \bf{WASP-25b:}\\
    2018 Mar 29 & EFOSC2 - Grism \#11,  & 1.56-1.0-2.08 & 160  & 170 & 97/225  & 13:01:12.763, -27:30:48.59 (1)  & 0.964 \\ \hline 
    2018 Jun 20  & IMACS - 300mm @ 17.5$^{\circ}$ & 1.01-2.1 & 20 -- 40  & 203 &  684/1552  & 13:01:12.763, -27:30:48.59 (1) & 0.964\\
     &  &   &  &    & &  13:02:14.300, -27:48:06.60 (2) & 0.457\\
      &  &   &  &    & &  13:00:50.950, -27:44:27.90 (3) & 0.205\\
      &  &   &  &    & &  13:01:54.318, -27:42:21.97 (4) & 0.191\\ \hline
    2021 Mar 06  & IMACS - 300mm @ 17.5$^{\circ}$ & 1.54-1.0-1.08 & 30  & 321 & 1386/2093  & 13:01:54.322, -27:42:21.80 (4)  & 0.191  \\
     &  &   &  &   & & 13:02:40.089, -27:24:34.76 (5)  &  0.074  \\
      &  &   &  &   & & 13:02:01.341, -27:46:21.51 (6) &  0.15 \\  
      &  &   &  &   & & 13:01:40.804, -27:35:03.73 (7) &  0.051 \\ \hline
    2022 Mar 25  & IMACS - 300mm @ 17.5$^{\circ}$ & 1.17-1.0-1.36 & 30  & 330 & 1635/2228  &     Same as on 06.03.2021 & ----------\\ \ChangeRT{1.6pt}
    \bf{WASP-124b:}\\
    2018 Sep 15 & IMACS - 300mm @ 26.7$^{\circ}$ & 1.21-1.0-1.33 & 100  & 160 & 700/1382  & 22:11:30.327, -30:45:56.50 (1) & 0.373 \\
     &  &  &  &  & & 22:11:40.676, -30:46:02.05 (2) & 0.501 \\
     &  &  &  &  & & 22:11:35.478, -30:44:10.86 (3) & 0.091 \\
     &  &  &  &  & &  22:10:18.862, -30:47:15.11 (4) & 0.137 \\
       &  &  &  &  & & 22:12:17.072, -30:47:31.72 (5) & 0.113 \\ \hline
    2019 Jun 15  & IMACS - 300mm @ 26.7$^{\circ}$ & 1.14-1.0-1.05 & 80 -- 100  & 111 &  757/1387 &  Same as on 15.09.2018 & ---------- \\  \hline
    2019 Aug 26 & IMACS - 300mm @ 17.5$^{\circ}$ & 1.41-1.0-1.08 & 60 -- 150  & 111 & 613/1404  & 22:11:35.480, -30:44:10.84 (3) & 0.091\\
    &  &  &  &   & & 22:10:18.865, -30:47:15.10 (4) & 0.137\\
    &  &  &  &   & & 22:11:17.340, -30:32:38.37 (6) & 0.029  \\
     &  &  &  &   & &  22:10:07.432, -30:45:56.62 (7) & 0.099  \\ \hline
    2021 Aug 9  & IMACS - 300mm @ 17.5$^{\circ}$ & 1.65-1.0-1.05 & 60  & 217 & 813/1288  &  Same as on 26.08.2019  & ---------- \\  \hline 
    2021 Sep 5  & IMACS - 300mm @ 17.5$^{\circ}$ &  1.35-1.0-1.16 & 60  & 221 &  611/1236  &  Same as on 26.08.2019  & ---------- \\  \hline
    2021 Oct 2  & IMACS - 300mm @ 17.5$^{\circ}$ & 1.14-1.0-1.34 & 60  & 215  & 745/1952  &  Same as on 26.08.2019  & ---------- \\  \hline
    2022 Jun 5  & IMACS - 300mm @ 17.5$^{\circ}$ & 2.10-1.0 & 90  & 153 & 363/919 &  Same as on 26.08.2019  & ---------- \\  \hline
    \enddata 
 \tablenotetext{}{\textbf{Note:} The resolutions were calculated from the FWHM near the peak of the spectrum. Magnitudes used to calculate $D$ were obtained from the UCAC4 Catalog \citep{2013UCAC4}. The number in parenthesis represents the comparison label that refers to that specific star.}
\end{deluxetable*}

\section{Light curve Analysis} \label{sec:LC_analysis}
The general steps in our light curve analysis are the same as what has been implemented in previous ACCESS papers \citep[e.g.] []{McGruder:2022, Allen2022}. This includes first creating a photometric (white) light curve by combining all counts of the entire spectrum for each exposure. Systematics are removed from the white light curve and a transit is fit to obtain transit parameters. These parameters are then used to constrain the priors for fitting the spectrophotometric (binned) light curves. The binned light curves are produced by summing the light within a band of wavelengths. The appropriate widths and centering of bins were determined by considering spectrophotometric precision, the overlap of spectral bands from different observations, high telluric absorption regions, and the desire to properly probe for atmospheric features. The average bin widths were $\sim$ 150{\AA} for WASP-25b and 160 {\AA} for WASP-124b, with 90\AA{} bins centered on the Na and K doublets and wider bins on low throughput edges of the spectra. The final binning scheme used is demarcated with dotted vertical lines in Figure \ref{fig:ExtraSpec} and written for each bin in the first column of the Figures in Appendix \ref{Appx:LightCurves}.

\subsection{White light Curve Fitting} \label{WLC}
We detrended the white light curve (WLC) using a combination of Principal Component Analysis (PCA) and Gaussian Processes (GPs), which we refer to as \textit{PCA+GP} \footnote{Prior to the detrending process, we removed outliers via visual inspection after dividing the target light curve by the sum of the comparisons' light curves.}. This routine is identical to what was used by \cite{Yan2020, McGruder:2020, Weaver:2021, McGruder:2022}. It involves first performing singular value decomposition on a matrix composed of the comparison stars' light curves in magnitude space, which yield eigenvectors and values that allow us to identify principal components that capture features common to all comparison light curves. However, we still needed to reduce systematics unique to the target star, for which we use GPs. We used \texttt{george} \citep{Mackey2014_george} to construct and evaluate the likelihoods of a multidimensional squared-exponential kernel dependent on the auxiliary observables of airmass, FWHM, sky flux, trace, and wavelength solution drift. Details of the GP hyperparameter priors are shown in Table \ref{tab:WLC_priors}. We used nested sampling \citep[\texttt{PyMultiNest;}][]{2014BuchnerPyMultiNest} to explore the posteriors of applying each principle component, combined with the GP regression. \edit3{Lastly}, we performed Bayesian model averaging \citep[BMA;][]{Gibson2014_modelAvg} to combine those posteriors and produce the final detrended light curves.

The analytical transit model was produced using \texttt{batman} \citep{Kreidberg2015_batman}, where the priors on period (P), semi-major axis (relative to stellar radius, a/R$_s$), impact parameter (b), and time of mid-transit (t$_0$) were all normal distributions with means and standard deviations set by the values found in \citet[][Table 1]{McGruder:2023}. The priors for the radius of the planet relative to the star (R$_p$/R$_s$) and the parameterized quadratic limb darkening (LD) parameters q$_1$ and q$_2$ \citep{Kipping:2013} were set wider. Table \ref{tab:WLC_priors} also provides information on those and the other transit priors. 

\begin{figure*}[htb]
    \centering
    \includegraphics[width=1\textwidth]{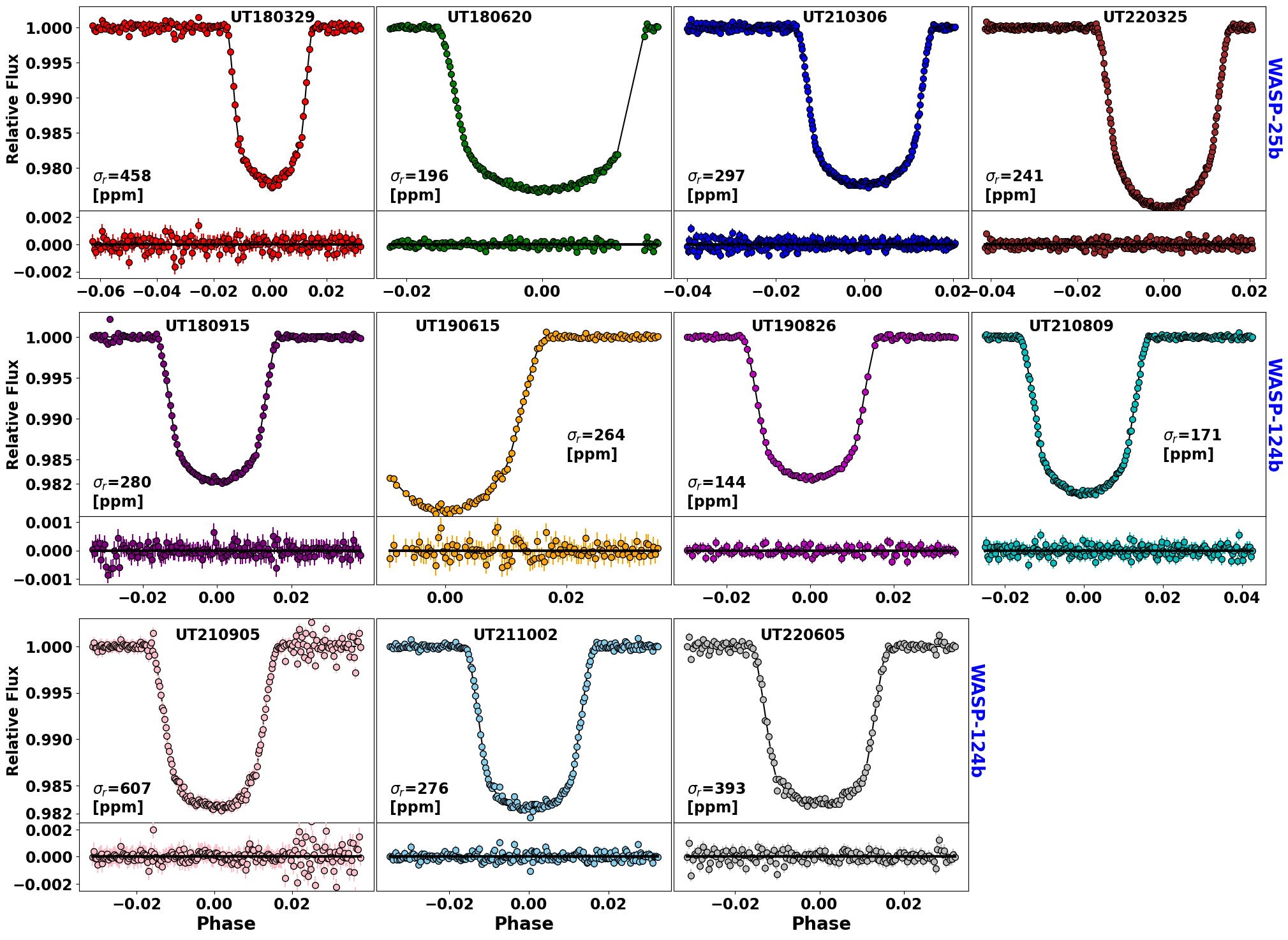}
    \caption{The final detrended white-light curves for each transit, utilizing the principal component analysis and Gaussian process detrending routine. The residuals obtained by the difference of the data from the best-fit model are shown below each light curve. The standard deviation of residuals are given by $\sigma_r$.}
    \label{fig:WLCs} 
\end{figure*}

After the transit parameters of the WLC were acquired for each night, we weight-averaged the P, a/R$_s$, and b values from each night to obtain more constrained terms for each. These means were held fixed for an additional pass of the PCA+GP run, to obtain final values for t$_0$, R$_p$/R$_s$, q$_1$ and q$_2$ \footnote{\edit3{We did not include a transit fitting pass with the common parameters free for the partial transit of UT190615 and just used the weighted means of the other six WASP-124b transits for fitting t$_0$, R$_p$/R$_s$, q$_1$ and q$_2$.}}. Figure \ref{fig:WLCs} shows the final detrended light curves of all transits and Table \ref{tab:wlc_PCAnGP} shows the values obtained when all parameters were fitted for (first four columns) and when the common parameters were fixed (last four columns). \edit2{From the table, one can see that only the LRG-BEASTS transit depth and the MOPSS partial transit depth differ by more than 2$\sigma$ between each other transit of a specific target. However, those two outliers are likely due to the LRG-BEASTS observations being bluer than those from ACCESS and the lack of full transit coverage from MOPSS hindering the detrending process.}

\begin{deluxetable*}{CC|c|c}[htb]
    \caption{White Light curve fitting priors}
    \label{tab:WLC_priors}
    \tablehead{&& \multicolumn{1}{|c|}{\bfseries {\large WASP-25b}} &\multicolumn{1}{|c|}{\bfseries {\large WASP-124b}}}
    \startdata 
    \textbf{parameter} & \textbf{function}  &\textbf{bounds}   &\textbf{bounds}  \\  \hline
    $\alpha$     & \text{log-uniform}       & 0.01--0.100 [ppm]        & 0.01--0.100 [ppm]  \\
    $\xi$   & \text{log-uniform}       & 0.01--0.100 [mmag]          & 0.01--0.100 [mmag] \\
    1/$\lambda$      & \text{gamma}        & $a$ = 1         & $a$ = 1 \\
    P               & \text{normal}       & m=3.7648337, $\sigma_n$= 1.2e-6        &  m=3.3726511, $\sigma_n$= 3.4e-6 \\
    t$_0$            & \text{normal}       &  m=2455274.99649, $\sigma_n$= 0.021    &  m=2457028.58329, $\sigma_n$= 0.021    \\
    $R_p/R_s$         & \text{normal}       &  m=0.139, $\sigma_n$= 0.02        &  m=0.125, $\sigma_n$= 0.02\\
    b                & \text{normal}         & m=0.357, $\sigma_n$= 0.042           & m=0.619, $\sigma_n$= 0.033\\ 
    $a/R_s$              & \text{normal}          & m=11.33, $\sigma_n$= 0.14           & m=9.22, $\sigma_n$= 0.13\\ 
    q$_1$                    & \text{uniform}          & 0--1            & 0--1\\
    q$_2$              & \text{uniform}          & 0--1           & 0--1\\ 
    \enddata 
    \tablenotetext{}{\textbf{Note:} The priors for the GP hyperparameters are amplitude ($\alpha$), jitter ($\xi$), and inverse squared length scale (1/$\lambda$). For 1/$\lambda$, when the $a$ parameter of a gamma function is set to 1, it becomes an exponential function (i.e. $e^{-x}$). The mean and standard deviation ($\sigma_n$) values of the transit parameters (variables defined in Section \ref{WLC}) were obtained directly from \cite{McGruder:2023}. With the exception of t$_0$, which was deduced for a given night from the period and t$_0$ obtained from \cite{McGruder:2023}. The $\sigma_n$ for this parameter was set to 30 minutes for each night.}
\end{deluxetable*}

\begin{deluxetable*}{|c|R|R|R|R|R|R|R|R|}[htb]
    \tabletypesize{\scriptsize}
    \caption{Fitted white light curve values} 
    \label{tab:wlc_PCAnGP}
    \tablehead{\colhead{Transit} & \colhead{P [days]} & \colhead{b} & \colhead{a/R$_s$} & \colhead{i [deg.]} & \colhead{R$_p$/R$_s$} & \colhead{t$_0$ (-2450000) [d]} & \colhead{q$_1$} & \colhead{q$_2$}}  
    \startdata 
    \multicolumn{9}{|c|}{\bf{WASP-25b:}}\\ \hline
     UT180329 & 3.7648337$^{\pm1.2e-6}$ & 0.360$^{+0.025}_{-0.026}$ & 11.28$^{\pm0.10}$ & 88.17$^{+0.15}_{-0.14}$ & 0.1343$^{+0.0017}_{-0.0019}$ & 8207.29525$^{+0.00025}_{-0.00026}$ & 0.523$^{+0.096}_{-0.090}$ & 0.36$^{+0.12}_{-0.11}$   \\ \hline 
     UT180620 & 3.7648337$^{\pm1.2e-6}$ & 0.329$^{+0.023}_{-0.025}$ & 11.199$^{+0.081}_{-0.075}$ & 88.32$^{+0.14}_{-0.13}$ & 0.1403$^{+0.0013}_{-0.0014}$ & 8290.62850$^{+0.00016}_{-0.00017}$ & 0.352$^{+0.066}_{-0.058}$ & 0.35$^{+0.12}_{-0.11}$  \\ \hline 
     UT210306 & 3.7648337$^{\pm1.2e-6}$ & 0.341$^{+0.022}_{-0.025}$ & 11.193$^{+0.075}_{-0.077}$ & 88.25$^{+0.13}_{-0.12}$ & 0.1374$^{+0.0025}_{-0.0021}$ & 9280.77911$^{\pm0.00012}$ & 0.531$^{+0.055}_{-0.049}$ & 0.199$^{+0.088}_{-0.086}$\\ \hline 
     UT220325 & 3.7648337$^{+1.1e-6}_{-1.2e-6}$ & 0.326$^{+0.021}_{-0.023}$ & 11.252$^{+0.084}_{-0.083}$ & 88.34$^{+0.13}_{-0.12}$ & 0.1452$^{+0.0025}_{-0.0024}$ & 9664.79213$^{+0.00023}_{-0.00022}$ & 0.458$^{+0.105}_{-0.093}$ & 0.42$^{+0.14}_{-0.11}$  \\ \hline
     mean & 3.76483368$^{\pm5.9e-7}$ & 0.338$^{\pm0.012}$ & 11.223$^{\pm0.042}$ & 88.275$^{\pm0.065}$& --- & --- & --- & ---  \\ 
    \ChangeRT{1.6pt}
    \multicolumn{9}{|c|}{\bf{WASP-124b:}}\\ \hline
     UT180915 & 3.372651$^{\pm3.3e-6}$ & 0.6302$^{+0.0132}_{-0.0165}$ & 9.154$^{+0.094}_{-0.093}$ & 86.055$^{+0.130}_{-0.118}$ & 0.1289$^{+0.0030}_{-0.0028}$ & 8377.64384$^{+0.00021}_{-0.00020}$ & 0.312$^{+0.062}_{-0.054}$ & 0.534$^{+0.217}_{-0.216}$  \\ \hline 
     UT190615 & --- & --- & --- & ---& 0.1399$^{+0.0038}_{-0.0039}$ & 8650.83074$^{+0.00076}_{-0.00079}$ & 0.478$^{+0.203}_{-0.139}$ & 0.655$^{+0.236}_{-0.304}$ \\ \hline 
     UT190826 & 3.3726511$^{+3.2e-6}_{-3.3e-6}$ & 0.6267$^{+0.0167}_{-0.0185}$ & 9.130$^{+0.109}_{-0.105}$ & 86.067$^{+0.148}_{-0.140}$   & 0.1265$^{+0.0060}_{-0.0071}$ & 8721.65346$^{+0.00043}_{-0.00052}$ & 0.450$^{+0.161}_{-0.145}$ & 0.277$^{+0.242}_{-0.169}$   \\ \hline 
     UT210809 & 3.3726509$^{+3.4e-6}_{-3.3e-6}$ & 0.6481$^{+0.0096}_{-0.0102}$ & 9.102$^{+0.082}_{-0.078}$ & 85.918$^{+0.092}_{-0.090}$  & 0.1333$^{+0.0026}_{-0.0025}$ & 9436.65573$^{+0.00016}_{-0.00017}$ & 0.409$^{+0.055}_{-0.053}$ & 0.411$^{+0.173}_{-0.171}$   \\ \hline 
     UT210905 & 3.372651$^{\pm3.3e-6}$ & 0.6062$^{+0.0199}_{-0.0212}$ & 9.204$^{+0.104}_{-0.105}$ & 86.223$^{+0.155}_{-0.153}$  & 0.1273$^{+0.0026}_{-0.0023}$ & 9463.63663$^{+0.00038}_{-0.00040}$ & 0.332$^{+0.133}_{-0.101}$ & 0.405$^{+0.301}_{-0.251}$  \\ \hline 
     UT211002 & 3.372651$^{\pm3.3e-6}$ & 0.6370$^{+0.0133}_{-0.0148}$ & 9.162$^{+0.082}_{-0.081}$ & 86.014$^{+0.121}_{-0.113}$  & 0.1273$^{+0.0014}_{-0.0013}$ & 9490.61771$^{\pm0.00020}$ & 0.431$^{+0.075}_{-0.073}$ & 0.144$^{+0.127}_{-0.095}$   \\ \hline 
     UT220605 & 3.372651$^{+3.2e-6}_{-3.1e-6}$ & 0.6452$^{+0.0118}_{-0.0130}$ & 9.175$^{+0.090}_{-0.087}$ & 85.969$^{+0.112}_{-0.106}$ & 0.1258$^{+0.0019}_{-0.0017}$ & 9736.82136$^{+0.00024}_{-0.00023}$ & 0.315$^{+0.066}_{-0.056}$ & 0.270$^{+0.258}_{-0.180}$  \\ \hline 
     mean& 3.372651$^{\pm1.3e-6}$ & 0.6379$^{\pm0.0056}$ & 9.152$^{\pm0.037}$ & 86.011$^{\pm0.048}$ & --- & --- & --- & ---  \\ \hline 
    \enddata
    \tablenotetext{}{\textbf{Note:} P, b, a/R$_s$, and i are the values obtained when leaving the transit parameters free, and R$_p$/R$_s$, t$_0$, q$_1$ and q$_2$ are values obtained from the second pass where only those transit parameters were allowed to be free and all others were fixed on the weighted mean values. Each printed value of t$_0$ is subtracted by 2450000 days. Transit UT190615 does not have the first four parameters because we did not allow those parameter to be free for the partial transit and just used the weighted means of the other six WASP-124b transits for fitting t$_0$, R$_p$/R$_s$, q$_1$ and q$_2$.}
\end{deluxetable*}

\subsection{Spectrophotometric Light Curves} \label{Binned}
We used two separate detrending routines to reduce the binned light curves (BLC). Both routines were discussed and tested in \cite{McGruder:2022}. For all WASP-25b transits and WASP-124b transits on UT190826, UT210809, and UT211002 we used common-mode correction (CMC) followed by polynomial correction (Poly), \textit{CMC+Poly}. This method assumes that most of the systematics are captured when detrending the WLC (following the procedures of Section \ref{WLC}). As such it uses the quotient of the final detrended WLC and the "raw" light curve, produced from the normalized target divided by the sum of comparison star light curves, as a common-mode term. The raw light curve also had a moving average 4-sigma clipping, similarly to \edit2{what} was done in \cite{McGruder:2022}. Each BLC is then divided by this common mode term. We then apply polynomial regession models dependent on the auxiliary observables (i.e. airmass, fwhm, see Section \ref{WLC}), to remove any additional chromatic systematics unique to each bin. For each auxiliary observable we allowed the polynomial to go up to fourth order, aside for airmass, which only went up to second order because of the smooth correlation with it and the light curves. We tested all combinations of each auxiliary parameter and order polynomials \footnote{1875 combinations for 5 observables and up to fourth order for everything but airmass, which is up to second order.} using \href{https://docs.scipy.org/doc/scipy/reference/generated/scipy.optimize.minimize.html}{\texttt{scipy.optimize.minimize}} \citep{scipy2020}. We then ranked each combination by polynomial corrections based on a standardized sum of $\chi^2$, Bayesian Information Criterion (BIC), Akaike Information Criterion (AIC), and root mean squared of the residuals (rRMS), and took the highest ranked 100 models (lower sums) to do a final pass of fitting with \texttt{PyMultiNest}. The priors on polynomial coefficients were normal with mean set by the \texttt{scipy.optimize.minimize} fits and standard deviation of one. We used BMA to combine posteriors and arrive at our final binned transit depths. 

The detrending method used for transits \edit2{UT180915}, UT190615, and UT210905 (three WASP-124b transits) was PCA+GP, which is the same algorithm used for our WLC analysis (see section \ref{WLC}). The reasoning for using this routine instead of the CMC+Poly is because the systematics found in each bin are significantly different from one another (see the corresponding first columns of figures in Appendix \ref{Appx:LightCurves}), so the CMC term poorly corrects the bins and likely introduces more systematics, which simple polynomials have trouble modeling. In fact, the transits that required PCA+GP had obvious issues with their observations such as missing ingress, scattered cloud coverage, and poor seeing. However, we also tested the performances of both methods by examining the binned spectra each produced, in particular we compared the variance of their transmission spectra and the average residual red noise of each bin $\beta$ \citep[][see appendix D]{McGruder:2022}. We found that for the three aforementioned WASP-124b transits, $\beta$ and the variance were substantially worse. This supports a finding of \cite{McGruder:2022}, in which the appropriate detrending method should be considered by testing multiple approaches for each dataset.

For both fitting methods all transit parameters were fixed to the WLC best fit (columns 2--5 of Table \ref{tab:wlc_PCAnGP}), but the LD parameters were uniform from 0 to 1 and R$_p$/R$_s$ had a normal prior with mean set by the WLC best fits and a standard deviation of 0.02.

\section{Transmission Spectra} \label{sec:TransSpec}
We produced our transmission spectra by comparing the best fit R$_p$/R$_s$ found for each detrended wavelength bin. We then combined the transmission spectra from each night to form a global transmission spectrum for each target. The spectra from each night had an offset applied so the mean depths across overlapping spectral bins were the same, as was done by \cite{McGruder:2022}. The spectra were then combined by weight averaging each overlapping bin. The new points were obtained using the python \texttt{numpy.average}\citep{numpy2020} function where the weight of each R$_p$/R$_s$ was given by the inverse squared R$_p$/R$_s$ errors ("\textit{inverse variance weighting}"). The error of the resulting weighted R$_p$/R$_s$ were calculated as the square root of weighted variance (again using inverse squared R$_p$/R$_s$ error as weights). \edit2{For WASP-25b the overlapping bins were from 4570--7113{\AA} and had a mean depth of 0.14040 R$_p$/R$_s$, which corresponded to weighted offsets of $0.00584$, $0.00016$, $0.00265$, and $-0.00552$ R$_p$/R$_s$ for each night in chronological order. For WASP-124b the mean depth was 0.12811 R$_p$/R$_s$ where all wavelengths overlapped aside for 6317--6424 and 9100--9225\,{\AA}. This yielded offsets of $-0.00047$, $0.00122$, $-0.00494$, $0.00049$, $0.00093$, and $0.00295$, respectively.}

\begin{figure*}[htb]
    \centering
    \includegraphics[width=1\textwidth]{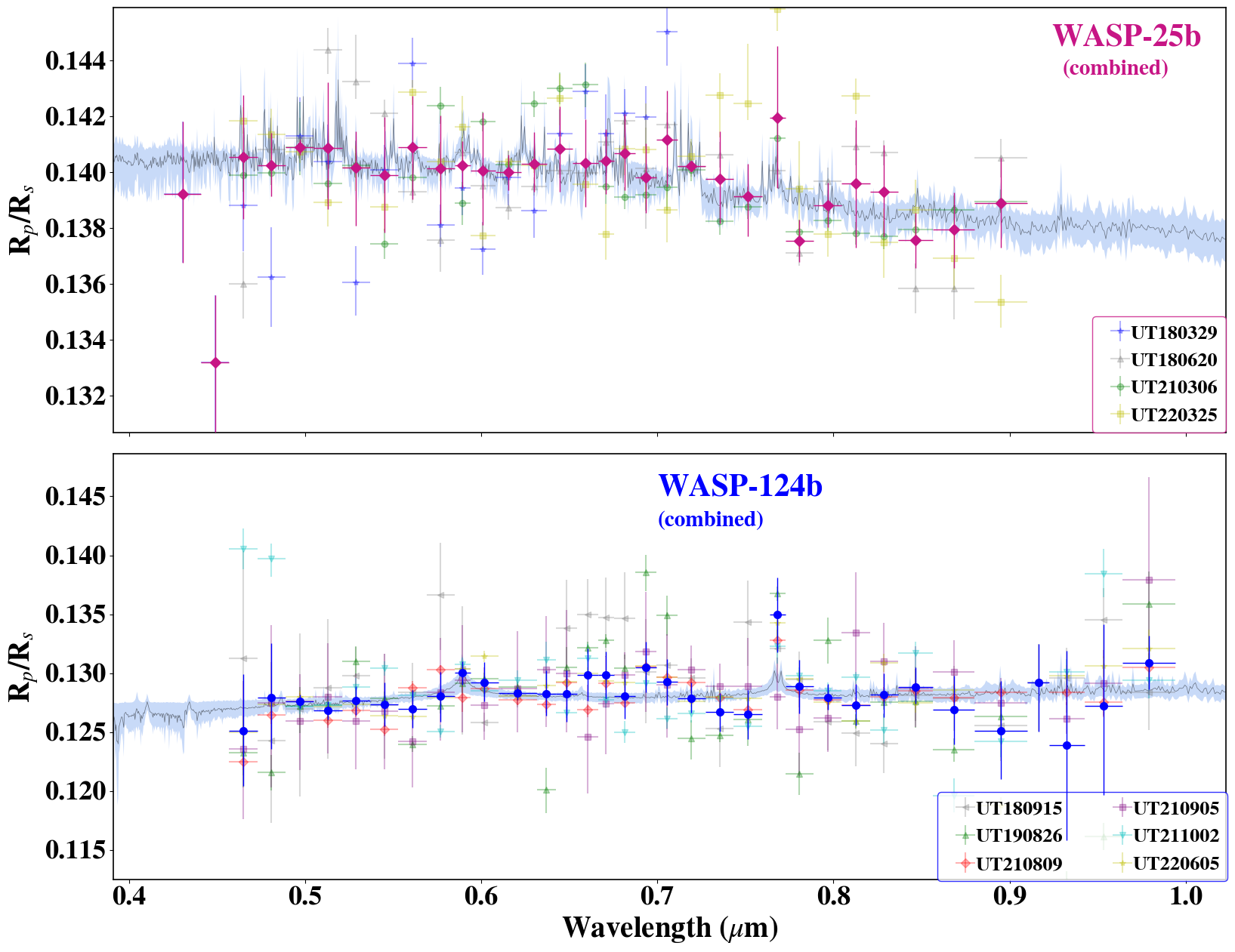}
    \caption{The final transmission spectra of WASP-25b (top) and WASP-124b (bottom). The final weighted-average spectra are shown in violet for WASP-25 and blue for WASP-124, with their individual transmission spectra used for the combined spectra plotted in transparent colors. The best fit \texttt{PLATON} retrieval models are plotted as a black line with the 1$\sigma$ confidence interval highlighted in light-blue. For both targets the best fit models are the ones that just include activity, but in the case of WASP-124b, this model is not significantly preferred over any other model.}
    \label{fig:TransSpecs} 
\end{figure*}

For WASP-124b, the partial transit on UT190615 was not included because the scatter of that transmission spectrum was too high to positively contribute to the combined spectrum. This is likely because the lack of ingress prevented both detrending methods (PCA+GP and CMC+Poly) from accurately constraining the systematics. Figure \ref{fig:TransSpecs} shows the transmission spectra of each night (except transit UT190615) and the combined transmission spectra. 

The average precision \edit2{(68\% confidence interval) of the combined spectra} per bin\footnote{The average bin size was $\sim$ 150{\AA} and 160{\AA} for WASP-25b and WASP-124b, respectively.} in R$_p$/R$_s$ for the WASP-25b data was 0.00301 (841 ppm in depth) and \edit3{0.00485 (1238} ppm in depth) for WASP-124b. With these precisions, we can only probe as low as 4.03 and \edit3{5.71} atmospheric pressure scale heights for WASP-25b and WASP-124b, respectively, for which the scale heights are 453.2~km and 634.9~km. Thus, it is difficult to determine if the relatively flat spectra seen in both targets (see Figure \ref{fig:TransSpecs}) is due to a true dearth of planetary features or a precision limitation. 

It should also be noted that even though there are more transits for WASP-124b, its transmission spectrum is less precise than the four transits of WASP-25b because WASP-124 (V$_{mag}$ = 12.7) is dimmer than WASP-25 (V$_{mag}$ = 11.9) and the overall quality of the WASP-124b observations were not as high, implied in section \ref{Binned}, where we apply the PCA+GP detrending routine for \edit2{2} out of the 6 used transits because of aggressive systematics.

\subsection{Retrieval Analysis} \label{sec:Retrievals}
We ran a series of retrieval models with \texttt{PLATON} \citep{Zhang_2019PLATON} and \texttt{Exoretrievals} \citep{Espinoza2019} on our final combined transmission spectra of WASP-25b and WASP-124b. Our analysis process was to run models including stellar activity, scattering features, or common atomic/molecular species observed in planets of this type (i.e., H$_2$O, Na, K) and the different combinations of each with \texttt{Exoretrievals}. Concurrently, we run models including scatters, stellar activity, neither, and both with \texttt{PLATON}, which assumes equilibrium chemistry and fits for the C/O ratio and planetary metallicity to extrapolate the abundances. This retrieval analysis workflow is the same as was done by \edit3{\cite{McGruder:2020} for WASP-31b}, \cite{McGruder:2022} for WASP-96b, and \cite{McGruder:2023} for WASP-6b and WASP-110b. To determine which models are preferred over another we used the Bayesian evidences (Z), given for each model because both retrieval routines use nested sampling to explore the posterior space (\texttt{PyMultiNest} with \texttt{Exoretrievals} and \texttt{dynesty} \citep{Speagle_2020Dynesty} with \texttt{PLATON}). Following the reasoning of \cite{2008Trotta} and \cite{2013Benneke}, we considered a $\Delta$lnZ less than 2.5 not significantly favoring one model over another, $\Delta$lnZ between 2.5 and 5 moderately favoring the higher evidence model, and greater than 5 strongly supports the higher lnZ model\footnote{A loose conversion of this to frequentist terms is $\Delta$lnZ of 2.5 $\sim$ 2.7$\sigma$ favoring and $\Delta$lnZ of 5 $\sim$ 3.6$\sigma$ favoring.}. \edit2{A table of the difference in natural-log evidences ($\Delta$lnZ) of a given model relative to the least complex model for the given retrieval is shown in Table \ref{tab:W25_W124_LnZs}. The term we use for when there are no scatters in the planetary atmosphere or activity in the stellar photosphere added to the model is \emph{plain}. Given that \texttt{PLATON} assumes equilibrium chemistry and the atomic/molecular abundances are inherently determined through this, a \emph{plain} model is the least complex model used for \texttt{PLATON}. In the case of \texttt{Exoretrievals}, its least complex model is when the presence of species is turned off, i.e. a \emph{plain} model (no scatters or activity), but also without atomic/molecular species included. In that case the spectrum is completely \emph{flat}.}


When determining the priors for the stellar activity parameter in our retrieval analysis, we used log$_{10}$($R'_{HK}$) and stellar rotational periods obtained from high resolution spectra and photometric observations in \citet[][Table 1]{McGruder:2023}. WASP-124 has a log$_{10}$($R'_{HK}$) of -4.765$\pm$0.056, consistent with WASP-96b's (log$_{10}$($R'_{HK}$ = -4.781$\pm$0.028), which has been established to be quiet \citep{Nikolov2022, McGruder:2022}. However, WASP-124 has a faster rotational period (10.65$^{+3.27}_{-3.01}$ days), which has been found to be correlated to activity levels \citep[e.g.,][]{Pizzolato2003,Wright2011,Wright2013}. With this in consideration we allow the covering fraction of unocculted inhomogeneities to vary uniformly from 0 to 6.8\%, which is consistent with the $2\sigma$ upper level of activity for stars of this type found by \citet[][see their Tables 2 \& 3 and Equation 2]{Rackham:2019}. For WASP-25, the log$_{10}$($R'_{HK}$) of -4.507$\pm$0.119 and faster stellar rotational period of 16.93$^{+2.02}_{-1.55}$ days suggests it is a somewhat active star; \edit2{as such, we do not limit its stellar inhomogeneity coverage and set the covering fraction priors to be uniform from 0 to 50\%. The priors used for each retrieval run are given in Appendix \ref{Appx:Retrievals}.}

\subsection{Retrieval Interpretation} \label{sec:RetrivInterp}
The best-fit parameters from the retrieval analysis can be seen in Table \ref{tab:BestRetrievedPar}. Overall, the \texttt{PLATON} and \texttt{Exoretrievals} results agree with each other well for each target. 
The lack of prominent features in either spectrum makes it difficult for planetary atmospheric properties to be constrained, which is outlined by the wide $1\sigma$ range given for every parameter in Table \ref{tab:BestRetrievedPar}. For WASP-25b, the best-fit models were \edit2{plain} atmospheres (i.e., no scatterers or atomic/molecular features) for the exoplanet and an inhomogeneous photosphere for the stellar host, with \edit3{$\sim$20\% coverage of cold spots at a temperature contrasts of $\Delta$T $\sim$ -2000 with respect to the quiescent photosphere. However, the uncertainty of these inhomogenity parameters is large (see Table \ref{tab:BestRetrievedPar}), with the most extreme case being the retrieved \texttt{PLATON} inhomogenity temperature of -2001$^{+1473}_{-531}$~K. For WASP-124b, using \texttt{Exoretrievals}, the highest-evidence models were those with low levels of stellar activity or a plain planetary atmosphere. Even still, those evidences were indistinguishable from a flat line model. The \texttt{PLATON} retrievals had the same issue where no model was significantly favored over another. This emphasizes the difficulty of constraining the atmosphere of WASP-124b with the data at hand.}

\begin{figure*}[htb]
    \centering
    \includegraphics[width=1\textwidth]{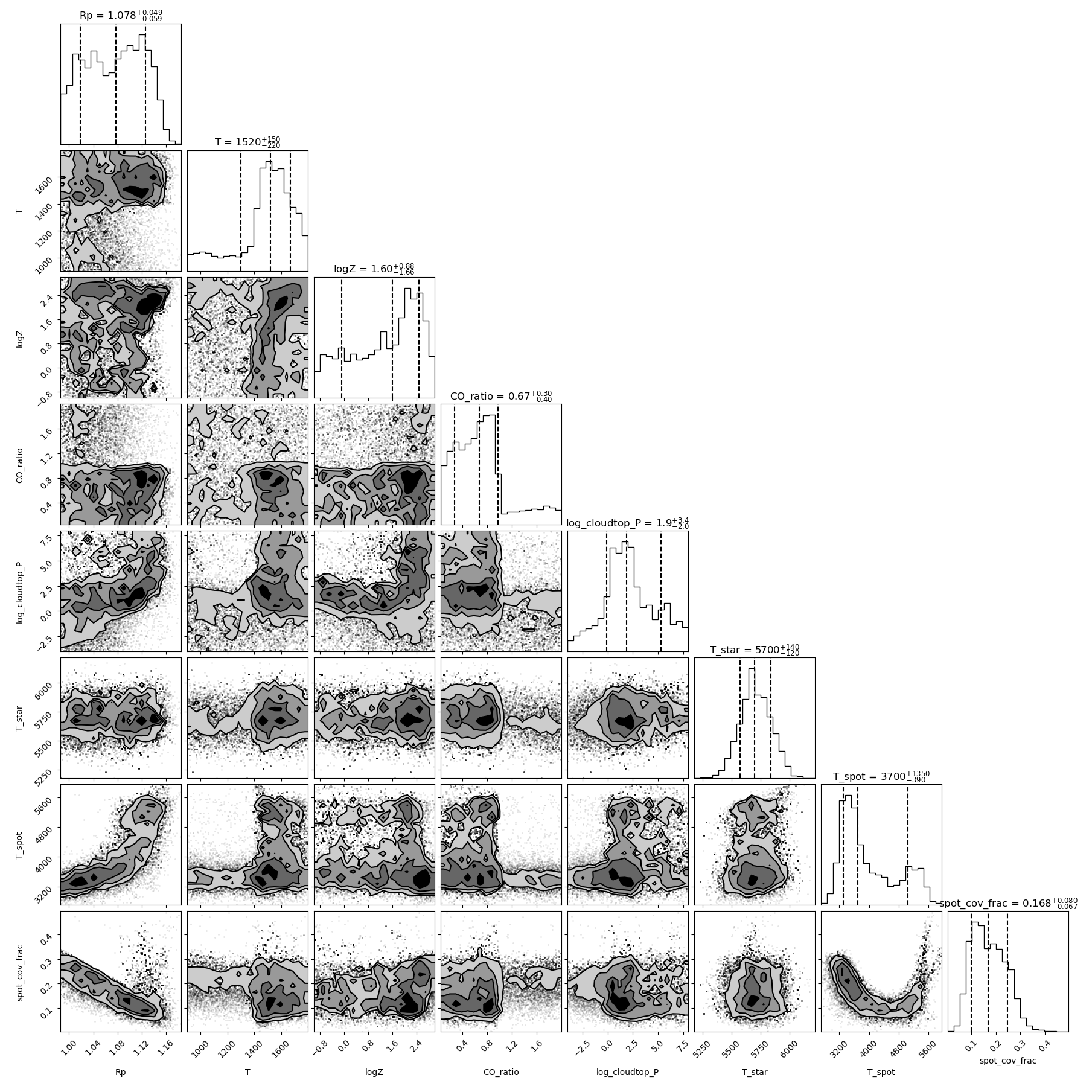}
    \caption{The corner plot obtained for the \texttt{PLATON} best fit of WASP-25. The best fit was one with activity and no additional scatters. However, it was only slightly favored over other models. As the posteriors outline, the lack of significant features makes it difficult to strongly retrieve properties of the planetary atmosphere.}
    \label{fig:CornerPlot} 
\end{figure*}

The limb temperatures obtained for WASP-25b and WASP-124b using both retrieval methods (see Table \ref{tab:BestRetrievedPar}) are in agreement with their corresponding effective temperatures of 1217$\pm$101K and 1481$\pm$123K, respectively \citep[][Table 1]{McGruder:2023}. However, again the uncertainties in the retrieved values are quite large. The pressures where the atmospheres are optically thick were also poorly constrained, with values ranging from log$_{10}$[bars] of -3.0$^{+3.8}_{-3.4}$ and -2.8$^{+3.7}_{-3.0}$ for WASP-25b and from -3.1$^{+3.4}_{-2.0}$ and -0.7$^{+2.2}_{-3.3}$ for WASP-124b, with \texttt{Exoretrievals} and PLATON, respectively. Thus pressures from $\sim$0.4 $\mu$ -- 0.2 bars are all within a 1-sigma interval for WASP-25b and from 1.6 $\mu$ -- 32 bars for WASP-124b. \edit2{Figure \ref{fig:CornerPlot}, shows the corner plot obtained for the \texttt{PLATON} best fit of WASP-25. It also highlights the difficulty in retrieving precise atmospheric parameters.}

We compared these results to analysis of WASP-31b \citep{McGruder:2020}, WASP-96b \citep{McGruder:2022}, WASP-6b, and WASP-110b \citep{McGruder:2023}. These planets were chosen because they underwent the same retrieval analysis, minimizing differences that may arise from varying model assumptions and priors \citep[e.g.,][]{Kirk:2019, Barstow:2020}\footnote{\cite{Barstow:2020} generally find consistency amongst the models and data tested, but do find cases where different models retrieve different parameters.}. 
The upper bounds of the pressure ranges for both targets are consistent with the 68\% interval of the WASP-96b fit, which strongly indicates the absence of aerosols in the observed wavelength range of 0.4--1.24~$\mu$m. However, their lower bounds are also consistent with the cloud top pressure found for WASP-110b, which has the highest retrieved cloud top \edit2{altitude (i.e. largest amount of high-altitude aerosols) } of the four planets. Thus, we reaffirm that further observations are needed to constrain the atmospheres of WASP-25b and WASP-124b. 

When attempting to interpret the relatively featureless spectra, we can deduce that it is unlikely that hazes are prominently present in the upper atmospheres of the planets, because there is no scattering slope observed. Strong scattering slopes in the optical have been suggested to signify hazes and could have signals as high as 15 scale heights \citep{Ohno2020}, well within the precision of the data. Therefore, the more likely cause for the observed features is either high-altitude clouds or observational limitations due to the lower precision, with one scenario not necessarily explaining both atmospheres. To obtain a better understanding of these atmospheres, higher precision optical observations with HST 
 and longer wavelength observations, ideally with JWST, are required. 

\begin{deluxetable*}{|l|C|C|C|C|C|C|C|C|C|C}[h!]
    \caption{$\Delta$ln~Z for \texttt{Exoretrievals} (left) and \texttt{PLATON} (right) retrievals}
    \label{tab:W25_W124_LnZs}
    \tablehead{& \multicolumn{7}{|c|}{\bfseries {\large Exoretrievals}} &\multicolumn{3}{|c|}{\bfseries {\large PLATON}}}
    \startdata 
      \textbf{Model:}  &  \text{flat}  &  $H_2O$  &  $K$  &  $Na$  &  $K +Na$  &  $H_2O +Na$ &  $H_2O + K +Na$  && \textbf{Model:} &\\ \hline
       \textbf{WASP-25b:} \\
      plain &	0.0 &	-0.81 &	-0.94 &	-1.01 &	-1.11 &	-1.0 &	-1.34  && \text{plain} & 0.0\\
      scattering &	--- &	-4.15 &	-3.91 &	-3.8 &	-3.78 &	-3.85 &	-3.94  && \text{scattering} & 2.89\\
      activity  &	5.0 &	\bf{6.28} &	6.25 &	6.16 &	5.81 &	5.85 &	5.61  && \text{activity} & \bf{4.13}\\
      Both &	--- &	-1.82 &	0.55 &	-0.35 &	-0.13 &	-2.07 &	-0.93 && \text{Both} & 3.75 \\ \hline
      \textbf{WASP-124b:} \\
    plain &	0.0 &	-0.83 &	-0.55 &	-0.34 &	-0.1 &	-0.35 &	-0.48 && \text{plain} & 0.0 \\
    scattering &	--- &	-3.17 &	-3.01 &	-3.12 &	-3.38 &	-3.41 &	-3.62 && \text{scattering} & -0.22 \\
    activity &	\bf{0.52} &	-0.95 &	-0.88 &	-0.62 &	-0.64 &	-0.88 &	-1.01 && \text{activity} & \bf{0.09} \\
    scattering+activity &	--- &	-3.42 &	-3.64 &	-3.63 &	-3.9 &	-3.93 &	-4.3 && \text{Both} & -0.27  \\
    \enddata 
    \tablenotetext{}{\textbf{Note:} The $\Delta$ln~Z values are relative to a plain (and flat for \texttt{Exoretrievals}'s case) spectrum with the combined WASP-25b (\textbf{top}) spectrum and the combined WASP-124b (\textbf{bottom}) spectrum. For WASP-25b the retrievals with activity were favored with both \texttt{Exoretrievals} and \texttt{PLATON}. Where as for WASP-124b no model had evidences strongly favoring it, but the plain models or the models with activity had higher evidences. The models with the highest evidences are highlighted in bold. In these models the ln~Z values for the flat (\texttt{Exoretrievals}, 0 $\Delta$ln~Z) and plain (\texttt{PLATON}, 0 $\Delta$ln~Z) models are -238 and 192 for WASP-25b, and -250 and 206 for WASP-124. We include those values for completeness, though the $\Delta$ln~Z is what is needed for model selection.}
\end{deluxetable*}

\begin{deluxetable*}{|l|C|C|l|C|C|}[h!]
    \caption{Parameters obtained by the best-fit retrievals for each system and retrieval code.}
    \label{tab:BestRetrievedPar}
    \tablehead{\multicolumn{3}{|c|}{\bfseries {\large Exoretrievals}} &\multicolumn{3}{|c|}{\bfseries {\large PLATON}}}
    \startdata 
          &  \textit{WASP-25b}  &  \textit{WASP-124b}  && \textit{WASP-25b}  &  \textit{WASP-124b}\\ \hline
     T\textsubscript{p} & 1350$^{+300}_{-290}$ & 1160$^{+430}_{-390}$ &  T\textsubscript{p}  & 1520$^{+150}_{-220}$  &  1090$^{+240}_{-160}$\\ \hline
     $\log_{10}(P\textsubscript{0})$  & -3.0$^{+3.8}_{-3.4}$ &  -2.8$^{+3.7}_{-3.0}$ & $\log_{10}(P\textsubscript{0})$   & -3.1$^{+3.4}_{-2.0}$ & -0.7$^{+2.2}_{-3.3}$  \\ \hline
      $\log_{10}(K)$   & -17.1$^{+9.0}_{-8.5}$ &  -12.3$^{+8.0}_{-12.0}$  &     &  &   \\ \hline
    $\log_{10}(Na)$   & -18.4$^{+8.5}_{-7.6}$ &  -9.7$^{+6.8}_{-14.1}$   &   $\log_{10}(Z/Z_{\odot}$)  & 1.6$^{+0.88}_{-1.66}$  &  1.2$^{+1.0}_{-1.3}$ \\ \hline
    $\Delta$T$_{het}$   & -2382$^{+372}_{-350}$ & 810$^{+1280}_{-1790}$    &  $\Delta$T$_{het}$ & -2001$^{+1473}_{-531}$    &  820$^{+1090}_{-1830}$   \\ \hline
    f$_{het}$   & 0.249$^{+0.057}_{-0.067}$  & 0.023$^{+0.028}_{-0.015}$    &  f$_{het}$ &  0.168$^{+0.08}_{-0.067}$   &  0.025$^{+0.026}_{-0.016}$  \\ \hline
    \enddata 
    \tablenotetext{}{\textbf{Note:} 
    For WASP-124b the heterogeneity parameters with \texttt{Exoretrievals} were obtained using the model that only included activity and the other parameters were obtained using the plain model that included K and Na. According to the evidences, both of those models were indistinguishable from each other (see Table \ref{tab:W25_W124_LnZs}). We used the model with activity, sodium, and potassium to obtain best fit parameters for WASP-25b. This model was used because all models including activity were indistinguishable from each other and this model obtained elemental mixing ratios. The difference in obtained overlapping parameters (i.e. T\textsubscript{p}, $\log_{10}(P\textsubscript{0})$, $\Delta$T$_{het}$, f$_{het}$) with that model and the one with just activity and water, were well within their uncertainties. The obtained water abundance was not constrained given there are no water features in the data, so we do not show its best fit here. Given that there are no carbon or oxygen bearing species in the wavelength coverage of our data, we do not report the C/O ratios retrieved by \texttt{PLATON}. The pressure in log$_{10}$(P$_0$) is given in bars for both retrievals.}
\end{deluxetable*}

\section{Similar Seven} \label{sec:Sim7}
\cite{McGruder:2023} proposed that there is a tentative trend with observed high-altitude aerosols and the host star metallicity for a group of seven planets with very similar system properties, which includes WASP-25b and WASP-124b. They claim if this trend is true, then WASP-25b and WASP-124b would sit on opposite ends, where WASP-25b would be obscured by aerosols (like WASP-6b and WASP-110b), and WASP-124b would be relatively clear (like WASP-96b).

We explore how well this trend holds here, using the sodium signal as a proxy for aerosol levels, as was done in \cite{McGruder:2023}. We find a small hint of Na in the spectrum of WASP-124b, which is stronger than for WASP-25b, even though the data precision of WASP-25b probes deeper. However, when looking at the retrieval analysis results of the WASP-124b data (see Table \ref{tab:W25_W124_LnZs}), we see no strong favoring of the \texttt{Exoretrievals} model which includes Na. Furthermore, the log mixing ratio of Na found with the \texttt{Exoretrievals} fit that included it is not well constrained and suggests marginal amounts of Na (-9.7$^{+6.8}_{-14.1}$). As such, we have no detection of Na in WASP-124b.

\begin{figure*}[htb]
    \centering
    \includegraphics[width=1\textwidth]{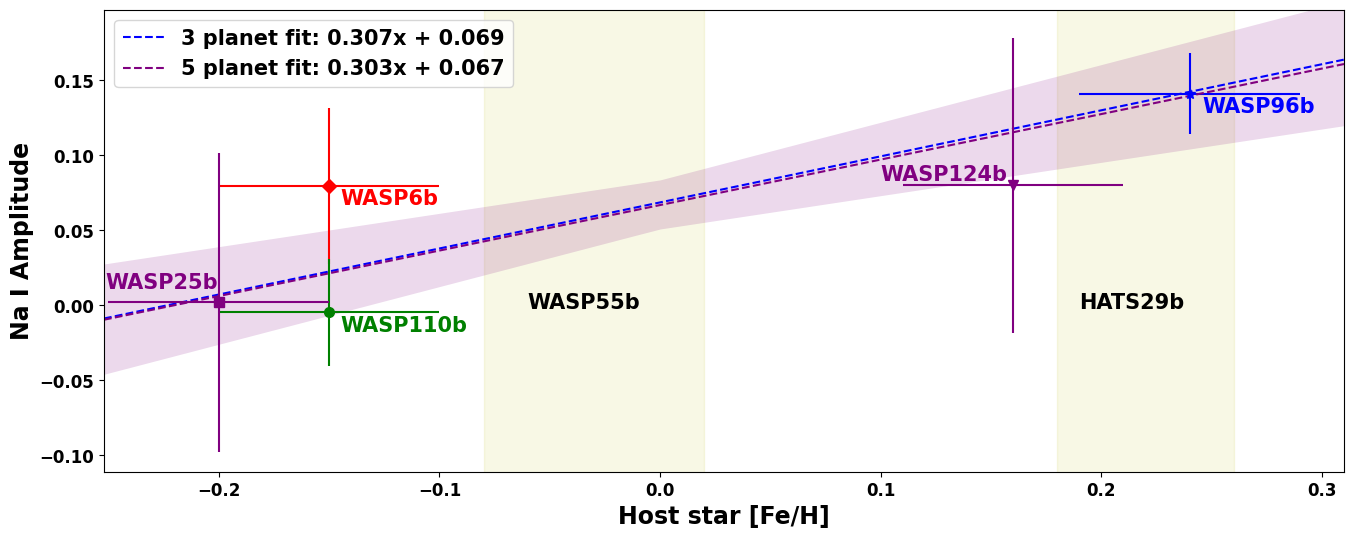}
    \caption{Sodium amplitude from the transmission spectra of WASP-6b \citep[red diamond][]{Nikolov:2015,Carter2020}, WASP-96b \citep[blue star][]{Nikolov:2018,McGruder:2022}, and WASP-110b \citep[green circle][]{Nikolov2021} versus their host star metallicities. In \cite{McGruder:2023}, "Na I Amplitude" "was the depth of the bin centered around the Na feature minus the average depth of all bins in the surrounding continuum. Here we take that value and divide by the theoretical difference of the peak depth from Na and the continuum, in order to incorporate slight differences in the planets' scale heights and temperature. \texttt{scipy.odr} was used to obtain a weighted linear fit with these three planetary signals and is shown as a dashed blue line. Its regression score, R$^2$, is 0.75. The 1$\sigma$ interval of the fit is shaded around the dashed line, and was calculated including the uncertainties in both metallicity and Na signal. Using the same method for WASP-6b and WASP-110b (i.e., R$_p$/R$_s$ bin centered at 5892.9\AA{} minus average R$_p$/R$_s$ within 5340--5820\AA{} and 5960--6440\AA), we calculated the Na signals of WASP-25b (purple square) and WASP-124b (purple triangle). A weighted linear fit with all 5 planet signals is shown as a purple dash-dotted line, with a R$^2$ of 0.71. Although our new measurements have larger uncertainties they are consistent with the trend identified in \cite{McGruder:2023}. The metallicity range of WASP-55b and HATS-29b are plotted as yellow shaded regions. } 
    \label{fig:CloudCorrelations} 
\end{figure*}

This seems inconsistent with the proposed trend, but the scale heights probed with the WASP-124b data is \edit3{5.71}. This is three times higher than what was able to be probed with WASP-6b, WASP-96b, and WASP-110b (2.004, 1.998, 2.212, respectively), which were used to identify the tentative trend. The precision of the other targets are likely higher because they include observations from larger (VLT) or space-based (HST) telescopes. In Figure \ref{fig:CloudCorrelations} we plot a linear aerosol--metallicity trend, where the Na feature was used as a proxy for aerosols, similar to what was done by \cite{McGruder:2023}. However, here we divide the Na amplitude values by their theoretical Na signal when no aerosols are present in the atmosphere, $\Delta$R$_p$/R$_s$ \citep[see equation 10 of][]{Heng:2016}. This was done, because though the planets are twin like, they are not exactly the same and would have slightly different maximum possible signals. Yet, because of the planets' similarity, this modification had little effect on the previous trend found by \cite{McGruder:2023}, as shown in Figure \ref{fig:CloudCorrelations}.  

In Figure \ref{fig:CloudCorrelations}, we plot a linear fit with and without the WASP-25b and WASP-124b Na signals, and find that given the errors both WASP-25b and WASP-124b are consistent with the trend found using just WASP-6b, WASP-96b, and WASP-110b. In Figure \ref{fig:CloudCorrelations} our linear trend is fit with \edit3{\texttt{scipy.odr}\citep{scipy2020}} where we used the inverse of each Na signals' errors as weights. The regression score, R$^2$, of the weighted linear fit with the Na signals from WASP-6b, WASP-96b, and WASP-110b is \edit3{0.755}. When including the Na signals from WASP-25b and WASP-124b, the score was \edit3{0.754}, showing that the trend continues to hold. We also compare the data to a flat line fit, i.e. no trend, and find a mean Na amplitude of 0.083 and R$^2$ of -0.051, emphasizing that a linear trend is much more appropriate given the data. 

Though there is no strong support for WASP-124b having high-altitude aerosols, if it does and is inconsistent with the tentative trend found by \cite{McGruder:2023}, a possible explanation might be due its difference in equilibrium temperature. All the planets' equilibrium temperatures lay around \edit3{$\sim$1250~K}, and though the equilibrium temperature of WASP-124b (1481$\pm$123~K) is less than 2-sigma from the coolest planet in our sample \edit2{(WASP-6b, T = 1167$\pm$96~K)}, it is possible that this temperature difference is important. Many studies find that equilibrium temperature is important in aerosol formation \citep[e.g.][]{Stevenson:2016, Fu:2017, Gao:2020, Estrela:2022}, but this literature is not consistent on if an increase in temperature for these class of planets would produce more or less aerosols. 
Thus, there is no strong support in the literature that the $\sim$300~K difference in equilibrium temperature of WASP-124b is an important factor in the tentative trend found. Still, if metallicity does not strongly correlate with aerosol formation rates, then perhaps an unobserved parameter, such as high energy emissions, is correlated to the higher aerosol rates observed in some of the \textit{Similar Seven} planets. Alternatively, the complex nature of aerosol formations in their extreme environments might make it difficult to correlate one or two parameters to the observed aerosol rates. 

To have a better grasp of if the aerosol--metallicity trend truly holds, at minimum, more optical observations of WASP-124b are needed to precisely probe its atmosphere. This is achievable with HST and/or larger ground-based telescope observations. Further Magellan/IMACS observations would also improve precision, especially if such observation were of good quality (i.e. the full transit, sufficient baselines, and good night conditions).

\section{Summary and Conclusions}\label{sec:Sum+Conc} 
 We observed four transits of WASP-25b, one with NTT/EFOSC2 and three with Magellan/IMACS, and seven transits of WASP-124b with Magellan/IMACS. We combined the transmission spectra from each night for each target (excluding the partial transit of WASP-124b on June 15th 2019) to produce near continuous final transmission spectra from 4200 -- 9100\AA{} for WASP-25b and from 4570 -- 9940\AA{} for WASP-124b. Our transmission spectra have an average precision in depth of 841~ppm for WASP-25b and \edit3{1238}~ppm for WASP-124b, corresponding to 4.03 and \edit3{5.71} scale heights, respectively. The spectrum of both targets lacked significant features. 
 
 Nevertheless, we ran a set of retrieval models utilizing \texttt{Exoretrievals} and \texttt{PLATON} on each final transmission spectrum. In doing so, \edit3{we found that the retrievals most favored model (with $\Delta \ln Z <$ 5) for WASP-25b is one that included $\sim$20\% covering fraction of unocculted cold spots at $\sim$2000~K cooler than the surrounding photosphere, but no molecular/atomic features. For WASP-124b there is no model strongly favored over another, but there are marginal hints of Na and K features and low levels of activity in the host star.} The retrieved atmospheric parameters from the best-fit models have wide uncertainties for both planets, but the retrieved limb temperatures are consistent with the calculated equilibrium temperatures. Given that there are no strong atomic or molecular features in either spectrum, the pressure levels where the atmosphere is optically thick and the atmospheric metallicities are poorly constrained. 
 The lack of features is possibly due to low precision being unable to probe the required depths for feature detection and/or high altitude clouds obscuring the spectra.

We then put these planets’ atmospheres in context with the aerosol--metallicity trend proposed by \cite{McGruder:2023}, and plot the sodium signal of these targets relative to their host stars’ metallicities. We find that the uncertainties of the Na signal, caused by lower data precision, makes it difficult to provide clear insight towards the existence of such a trend. We believe that further observations with higher quality data in the optical is necessary to confirm a trend. This could be done with HST and/or further ground-based telescopes. JWST observations would provide broader context of the nature of these planets atmospheres. Proving this trend has the potential to drastically direct theoretical understandings of aerosol formation and could yield more efficient target selection criteria.


\appendix
\section{Light Curves} \label{Appx:LightCurves}
\begin{figure*}[h!]
    \centering
    \includegraphics[width=.55\textwidth]{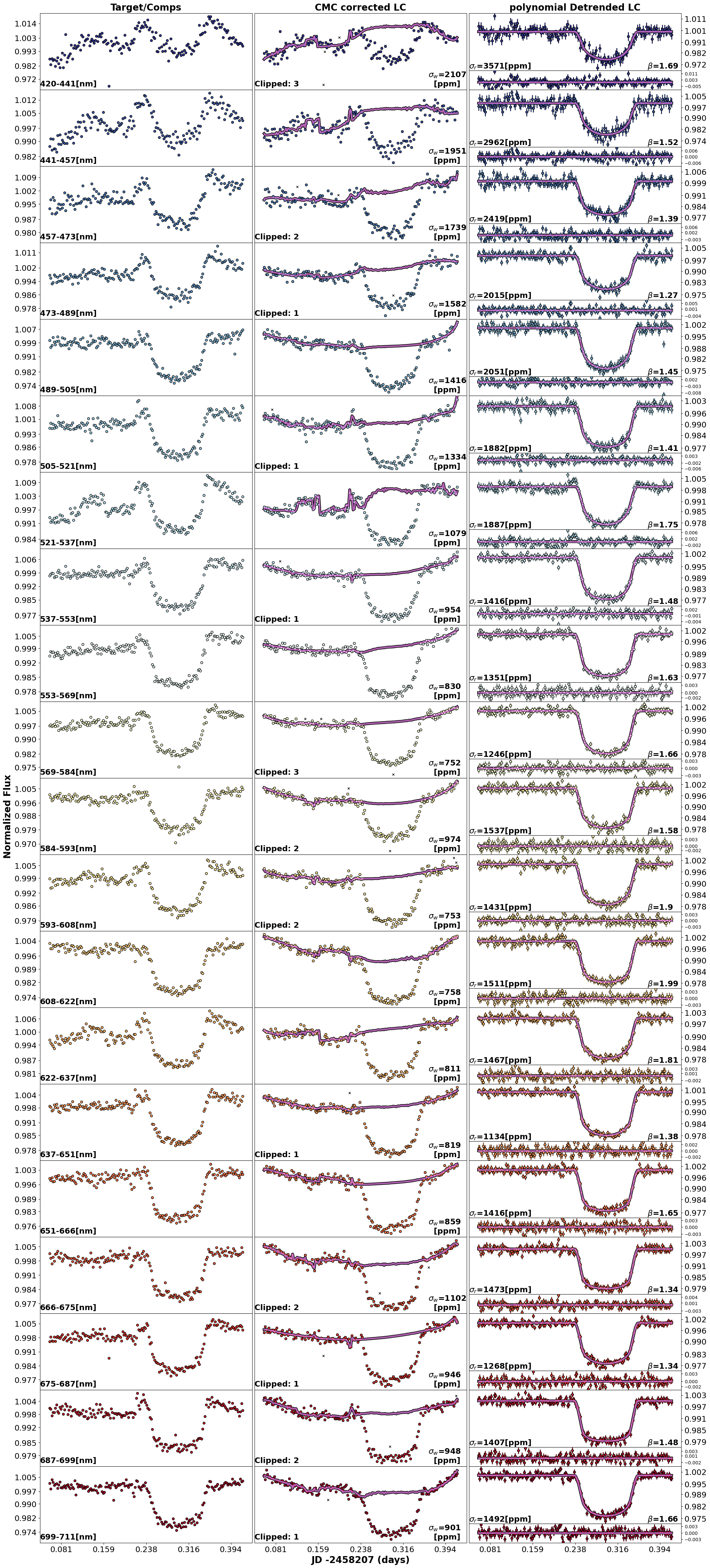}
    \caption{The light curves (LCs) in each step of the detrending process when utilizing the CMC+Poly method (see Section \ref{Binned}), for each bin of transit UT180329. \textbf{Left:} The LC produced from the quotient of the WASP-25 and comparison LCs. For all other transits there are multiple comparisons, in that situation the sum of comparison light curves are used instead. The wavelength range [nm] used is printed in the bottom left regions. \textbf{Middle:} The quotient of the LCs on the left and the common mode correction term. Overplotted in violet is the best fit polynomial correction term. If there were any frames clipped for a given bin, it is marked with a black x and the total number of clipped values is in the lower left regions. The precision, assuming only photon noise, is printed as $\sigma_{w}$. \textbf{Right:} The final detrended LC with the best-fit transit model overplotted in violet. The standard deviation of the residuals ($\sigma_{r}$) and $\sigma_{r}$/$\sigma_{w}$ ($\beta$) are printed. The error bars shown are determined from multiplying the photon noise precision of each frame by the $\beta$ for that bin.}
    \label{fig:LC_ut180329} 
\end{figure*}

\begin{figure*}[h!]
    \centering
    \includegraphics[width=.46\textwidth]{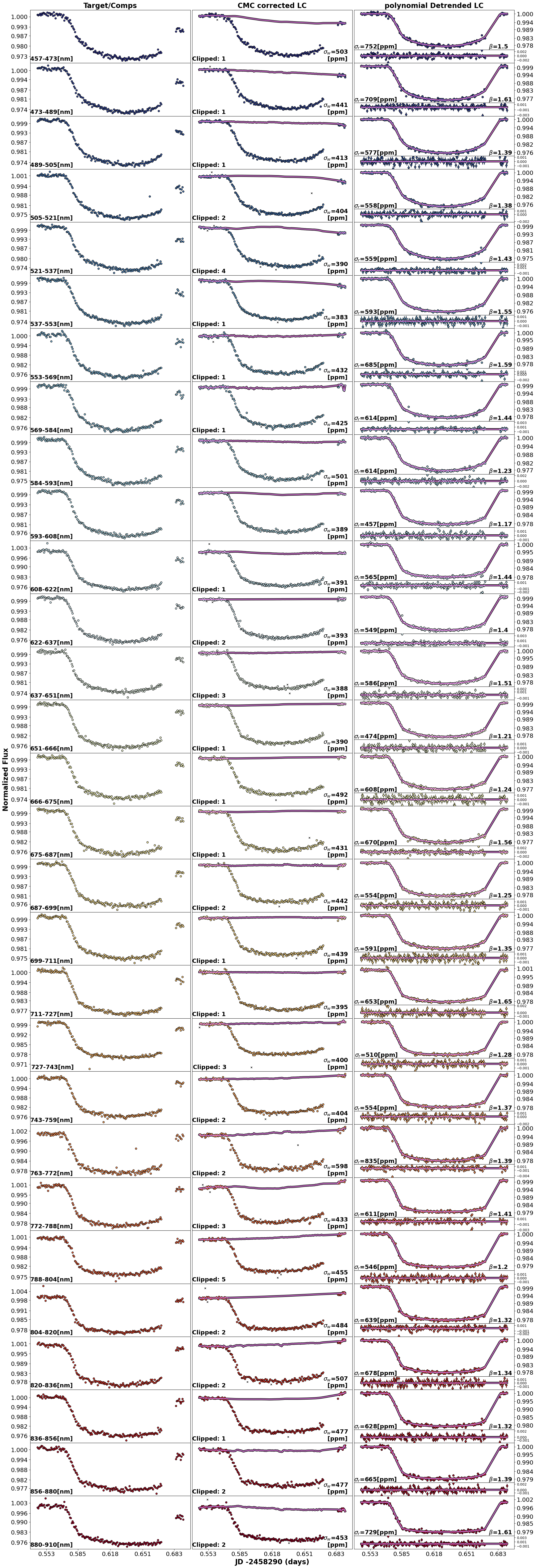}
    \caption{Same as Figure \ref{fig:LC_ut180329} but for transit UT180620.}
    \label{fig:LC_ut80620} 
\end{figure*}

\begin{figure*}[h!]
    \centering
    \includegraphics[width=.46\textwidth]{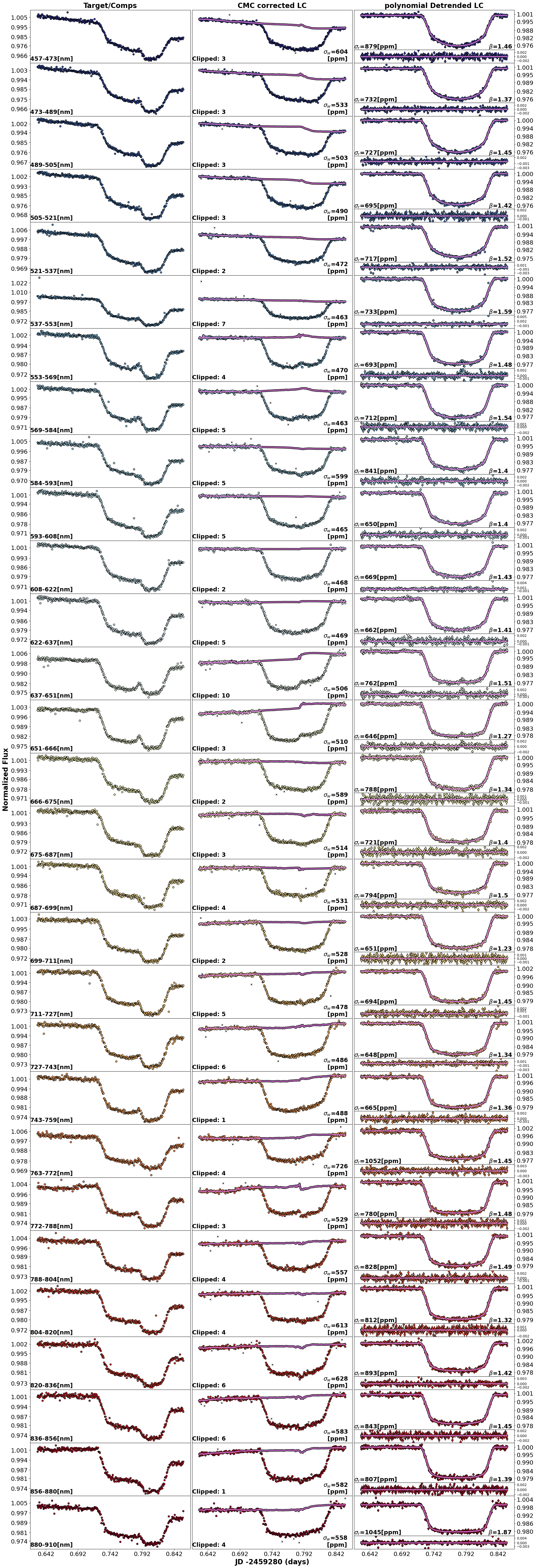}
    \includegraphics[width=.46\textwidth]{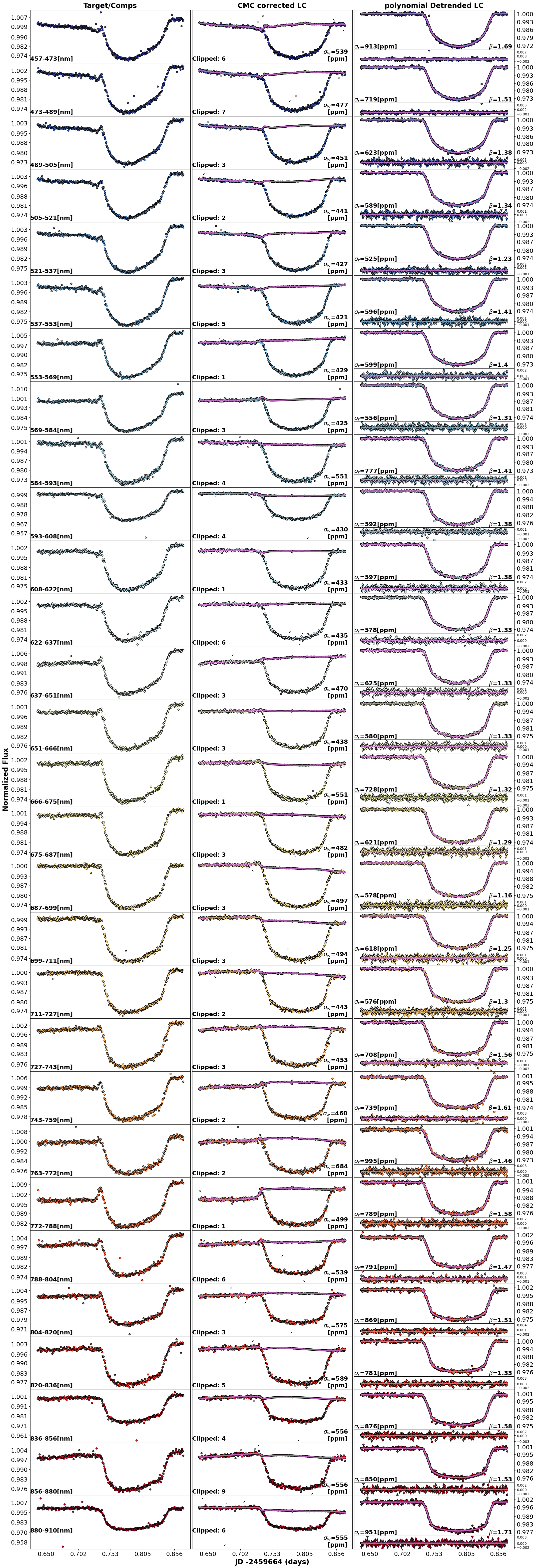}
    \caption{Same as Figure \ref{fig:LC_ut180329} but for transits UT210306 (\textbf{Left Figure}) and transit UT220325 (\textbf{Right Figure}).}
    \label{fig:LC_UT210306_UT220325} 
\end{figure*}

\begin{figure*}[h!]
    \centering
    \includegraphics[width=.34\textwidth]{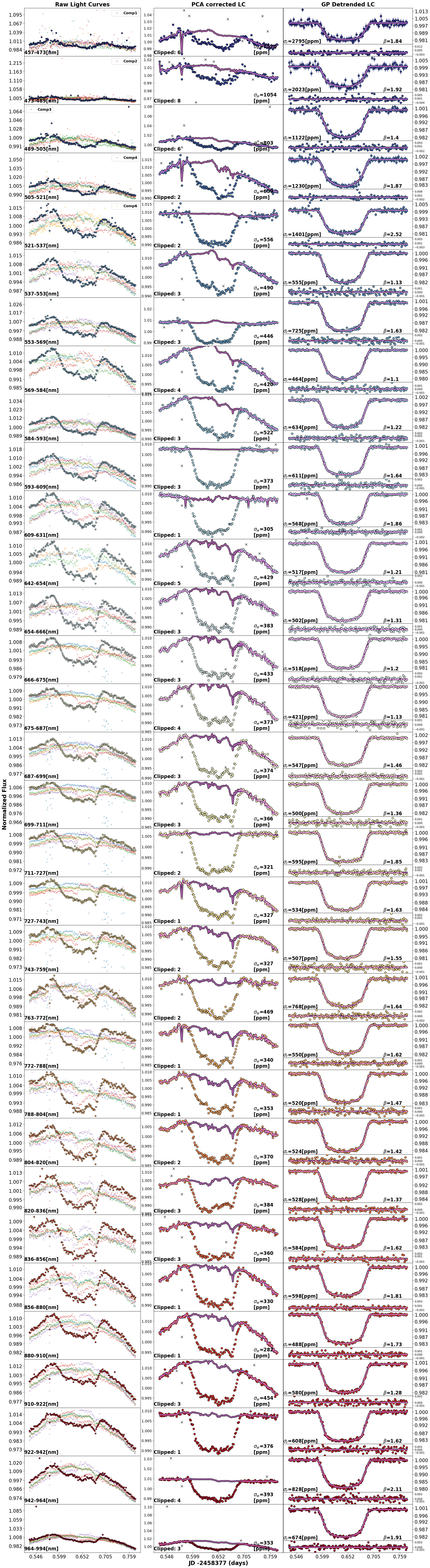}
    \includegraphics[width=.34\textwidth]{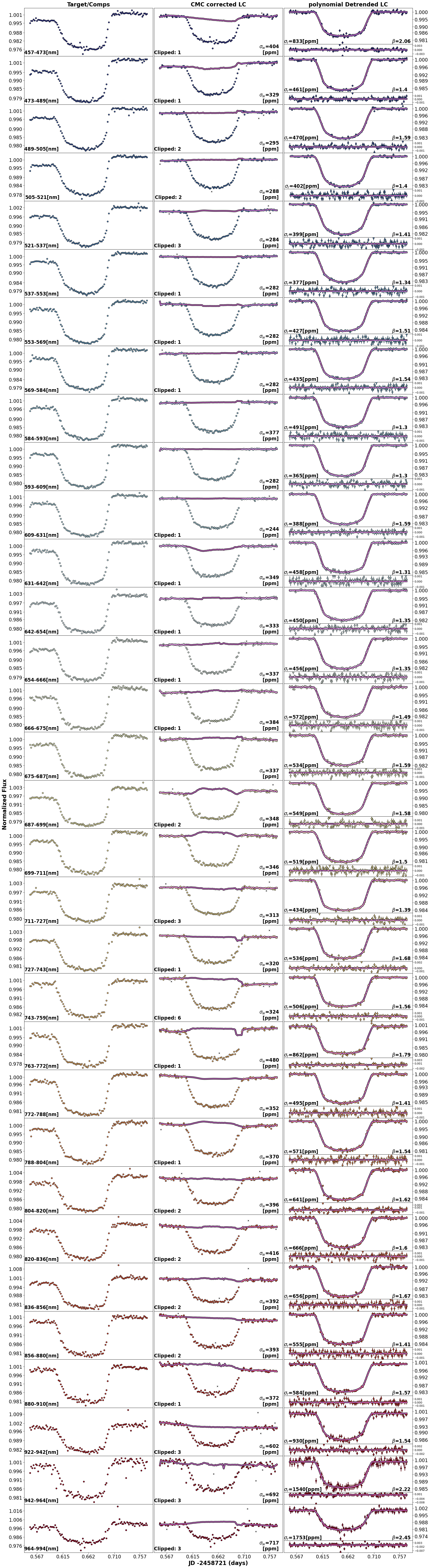}
    \caption{ \textbf{Left Figure:} The light curves (LCs) in each step of the detrending process when utilizing the PCA+GP method (see Sections \ref{sec:LC_analysis}), for each bin of transit UT180915. Each column shows, first on the left are the raw LCs produced directly from the extracted spectrum in Figure \ref{fig:ExtraSpec}, then in the middle the binned transit LCs of WASP-124b corrected with PCA (the best GP systematic fits are overplotted in violet), and on the right is the final detrended LC with the best fit transit model overplotted in violet. The best fit GP jitter term was used to approximate the error bars shown in the middle and right columns. The wavelength ranges used, clipped frames, photon noise precisions, standard deviation of residuals, and $\beta$ are printed on the figures just as in Figure \ref{fig:LC_ut180329}. \textbf{Right Figure:} Same as Figure \ref{fig:LC_ut180329} but for transit UT190826.}
    \label{fig:LC_ut180915_UT190826} 
\end{figure*}

\begin{figure*}[h!]
    \centering
    \includegraphics[width=.37\textwidth]{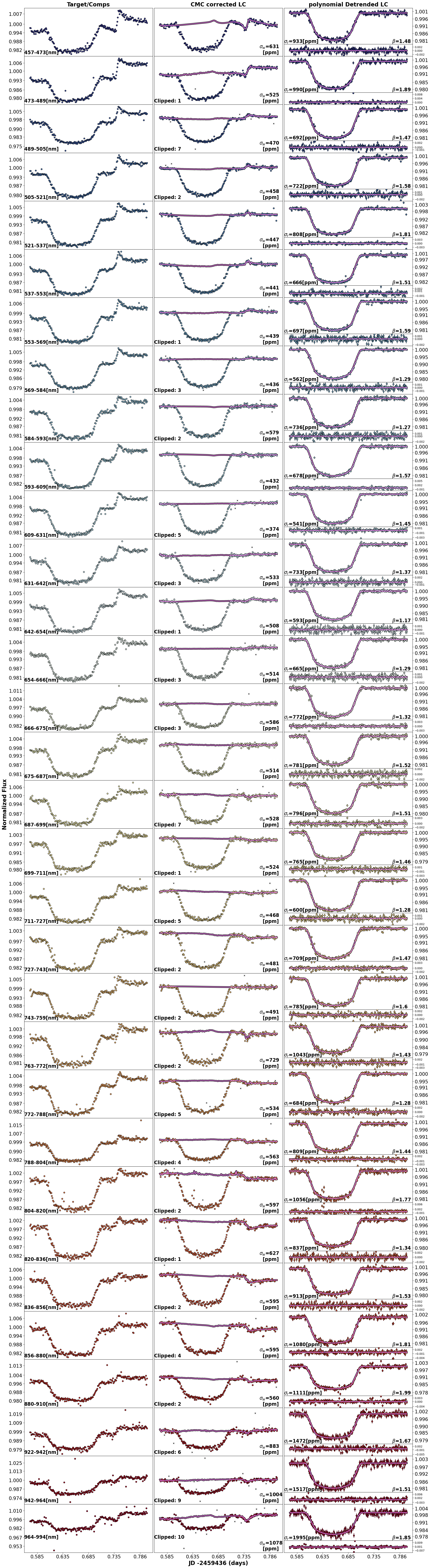}
    \includegraphics[width=.37\textwidth]{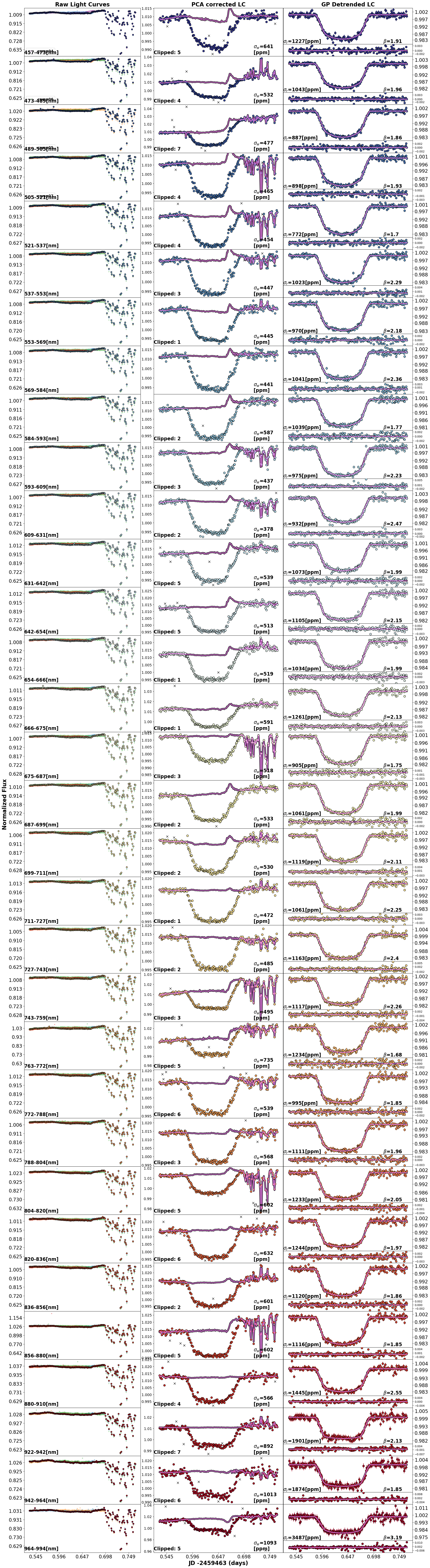}
    \caption{\textbf{Left Figure:} Same as Figure \ref{fig:LC_ut180329} but for transit UT210809. \textbf{Right Figure:} Same left side of Figure \ref{fig:LC_ut180915_UT190826} but for transit UT210905.}
    \label{fig:LC_UT210809_UT210905} 
\end{figure*}

\begin{figure*}[h!]
    \centering
    \includegraphics[width=.37\textwidth]{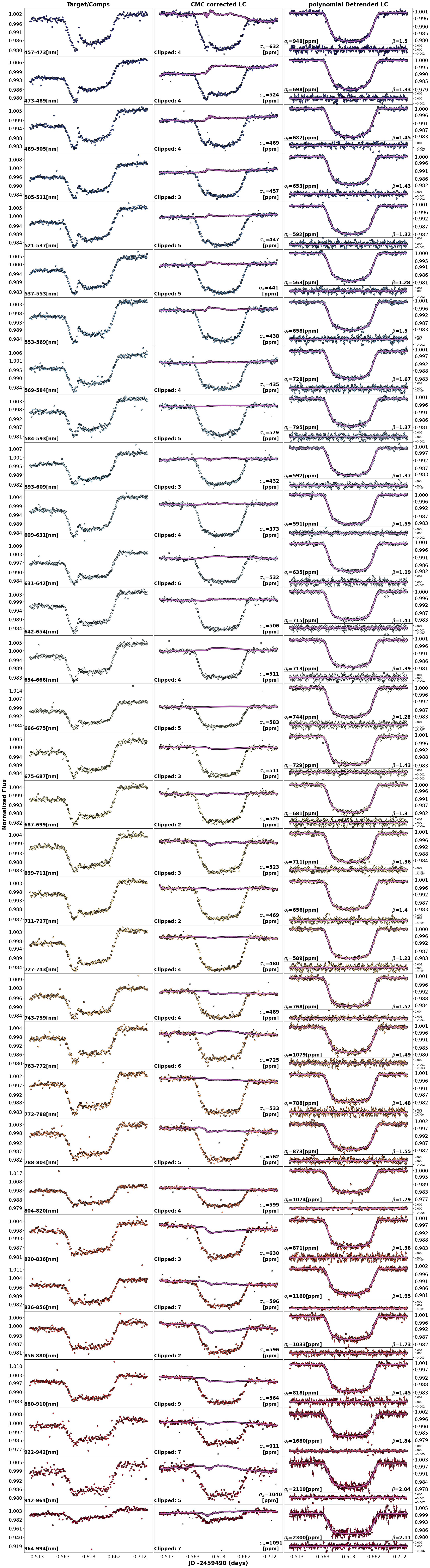}
    \includegraphics[width=.37\textwidth]{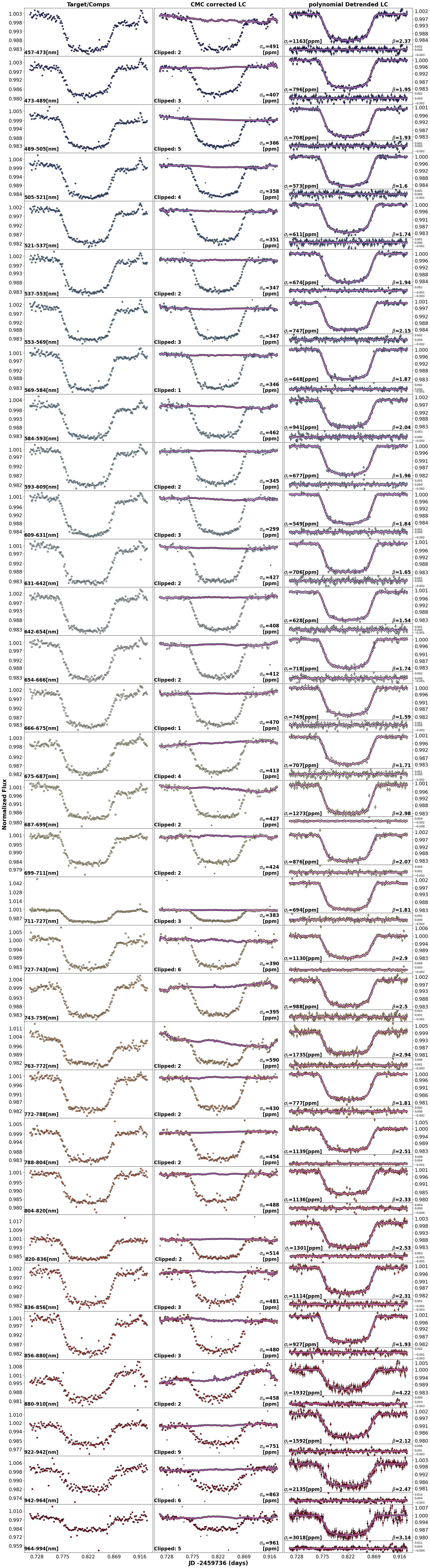}
    \caption{ Same as Figure \ref{fig:LC_ut180329} but for transit UT211002 (\textbf{Left Figure}) and transit UT220605 (\textbf{Right Figure}.)}
    \label{fig:LC_UT211002_UT220605} 
\end{figure*}

The detrending steps for the spectrophotometric bins are shown in Figures \ref{fig:LC_ut180329} -- \ref{fig:LC_UT211002_UT220605}, where all WASP-25b transits were detrended with the CMC+Poly detrending routine, transits \edit2{UT180915},
UT190615, UT210905, and UT220605 for WASP-124b were detrended with the PCA+GP detrending method, and the remaining WASP-124b transits were with CMC+Poly. Table \ref{tab:W25_W124_CombTransSpec} has the combined transmission spectra of WASP-25b (left) and WASP-124b (right). \edit2{Figures \ref{fig:LC_ut180329} -- \ref{fig:LC_UT211002_UT220605} and the transmission spectra of each individual night, including the unused partial transit of UT190615, can be obtained via \href{https://zenodo.org/record/8047731}{zenodo.org/record/8047731}.}

\begin{deluxetable*}{|C|C|c|C|C|}[htb]
    \caption{Transit depths (R$_p$/R$_s$) for the optical transmission spectra of WASP-25b and WASP-124b}
    \label{tab:W25_W124_CombTransSpec}
    \tablehead{\multicolumn{2}{|c|}{\bfseries {\large WASP-25b}} &&\multicolumn{2}{|c|}{\bfseries {\large WASP-124b}}}
    \startdata 
\textbf{Wavelength~({\AA})} & \textbf{R\textsubscript{p}/R\textsubscript{s}} & &\textbf{Wavelength~({\AA})} & \textbf{R\textsubscript{p}/R\textsubscript{s}}  \\  \hline
4200.0-4410.0 & 0.1392$^{+0.0025}_{-0.0026}$ & &  4570.0-4730.0 & 0.1251$^{+0.0047}_{-0.0048}$ \\ \hline
4410.0-4570.0 & 0.1332$^{+0.0025}_{-0.0024}$ & & 4730.0-4890.0 & 0.1279$^{+0.0043}_{-0.0047}$ \\ \hline
4570.0-4730.0 & 0.1405\pm0.0022 & & 4890.0-5050.0 & 0.1276\pm0.0006 \\ \hline
4730.0-4890.0 & 0.1402$^{+0.0011}_{-0.0012}$ & & 5050.0-5210.0 & 0.1269$^{+0.0008}_{-0.0007}$ \\ \hline
4890.0-5050.0 & 0.1409\pm0.0003 & & 5210.0-5370.0 & 0.1277$^{+0.0013}_{-0.0012}$ \\ \hline
5050.0-5210.0 & 0.1409$^{+0.0022}_{-0.0024}$ & &  5370.0-5530.0 & 0.1273\pm0.0018 \\ \hline
5210.0-5370.0 & 0.1402$^{+0.0021}_{-0.0013}$ & & 5530.0-5690.0 & 0.1269\pm0.0016 \\ \hline
5370.0-5530.0 & 0.1399\pm0.0021 & & 5690.0-5847.9 & 0.128$^{+0.0022}_{-0.0021}$ \\ \hline
5530.0-5690.0 & 0.1409$^{+0.0019}_{-0.0018}$ & & 5847.9-5937.9 & 0.13\pm0.001 \\ \hline
5690.0-5847.9 & 0.1401$^{+0.002}_{-0.0019}$ & & 5937.9-6097.9 & 0.1292$^{+0.0016}_{-0.0017}$ \\ \hline
5847.9-5937.9 & 0.1403\pm0.0008 & & 6097.9-6317.0 & 0.1283\pm0.0005 \\ \hline
5937.9-6082.9 & 0.1401$^{+0.002}_{-0.0021}$ & &  6317.0-6424.0 & 0.1283\pm0.0019 \\ \hline
6082.9-6227.9 & 0.14\pm0.0006 & &  6424.0-6542.86 & 0.1282$^{+0.0023}_{-0.002}$ \\ \hline
6227.9-6372.87 & 0.1403$^{+0.001}_{-0.0011}$ & &  6542.86-6662.86 & 0.1298$^{+0.0022}_{-0.0023}$ \\ \hline
6372.87-6517.86 & 0.1408\pm0.0015 & & 6662.86-6752.86 & 0.1299\pm0.002 \\ \hline
6517.86-6662.86 & 0.1403\pm0.0016 & & 6752.86-6872.86 & 0.1281\pm0.0019 \\ \hline
6662.86-6752.86 & 0.1404$^{+0.0012}_{-0.0011}$ & & 6872.86-6992.86 & 0.1305$^{+0.0021}_{-0.0022}$ \\ \hline
6752.86-6872.86 & 0.1407$^{+0.0013}_{-0.0014}$ & & 6992.86-7113.0 & 0.1293$^{+0.002}_{-0.0019}$ \\ \hline
6872.86-6992.86 & 0.1398$^{+0.0013}_{-0.0012}$ & & 7113.0-7273.0 & 0.1279$^{+0.0012}_{-0.0013}$ \\ \hline
6992.86-7113.0 & 0.1412\pm0.0017 & & 7273.0-7433.0 & 0.1267$^{+0.0017}_{-0.0016}$ \\ \hline
7113.0-7273.0 & 0.1402\pm0.0002 & & 7433.0-7597.0 & 0.1265\pm0.0021 \\ \hline
7273.0-7433.0 & 0.1398$^{+0.0016}_{-0.0017}$ & & 7636.5-7726.5 & 0.135$^{+0.0032}_{-0.0031}$ \\ \hline
7433.0-7597.0 & 0.1391$^{+0.0014}_{-0.0012}$ & & 7726.5-7886.5 & 0.1289\pm0.0023 \\ \hline
7636.5-7726.5 & 0.1419$^{+0.0025}_{-0.0026}$ & & 7886.5-8046.5 & 0.128\pm0.0011 \\ \hline
7726.5-7886.5 & 0.1376\pm0.0008 & & 8046.5-8206.5 & 0.1273$^{+0.0017}_{-0.0018}$ \\ \hline
7886.5-8046.5 & 0.1388\pm0.0008 & & 8206.5-8366.5 & 0.1282$^{+0.0019}_{-0.0018}$ \\ \hline
8046.5-8206.5 & 0.1396\pm0.0023 & & 8366.5-8566.0 & 0.1288$^{+0.0014}_{-0.0017}$ \\ \hline
8206.5-8366.5 & 0.1393$^{+0.0016}_{-0.0017}$ & & 8566.0-8800.0 & 0.1269\pm0.0029 \\ \hline
8366.5-8566.0 & 0.1376\pm0.001 & & 8800.0-9100.0 & 0.1251$^{+0.0041}_{-0.0043}$ \\ \hline
8566.0-8800.0 & 0.1379$^{+0.0014}_{-0.0013}$ & & 9100.0-9225.0 & 0.1292$^{+0.0042}_{-0.0033}$ \\ \hline
8800.0-9100.0 & 0.1389\pm0.0016 & &  9225.0-9425.0 & 0.1239$^{+0.0081}_{-0.008}$ \\ \hline
 &  & &  9425.0-9640.0 & 0.1272$^{+0.0076}_{-0.0069}$ \\ \hline
 &  & &  9640.0-9940.0 & 0.1309\pm0.0023 \\ \hline
    \enddata 
\tablenotetext{}{\textbf{Note:} Each spectrum was produced by weight averaging (with an offset) each of the transits for the given target. The transmission spectra for individual nights are uploaded on \href{https://zenodo.org/record/8047731}{zenodo.org/record/8047731}}
\end{deluxetable*}

\section{Atmospheric Retrieval Priors} \label{Appx:Retrievals}
Table \ref{tab:Priors4} has the priors used for each retrieval model.
\begin{deluxetable*}{|C|C|C|C|C|C|}[htb]
    \caption{The priors for \texttt{Exoretrievals} and \texttt{PLATON}}
    \label{tab:Priors4}
    \tablehead{\multicolumn{3}{|c|}{\bfseries {\large Exoretrievals}} &\multicolumn{3}{|c|}{\bfseries {\large PLATON}}}
    \startdata 
    \textbf{parameter}                          & \textbf{function} & \textbf{bounds} & \textbf{parameter}           & \textbf{function} & \textbf{bounds} \\ \hline
    \text{reference pressure (P\textsubscript{0}, bars)}                            & \text{log-uniform}      & \text{-8 to 3}        &  \text{reference pressure (P\textsubscript{clouds}, Pa)}           & \text{log-uniform} & \text{-3.99 to 7.99} \\ \hline
    \text{planetary atmospheric}                & \text{uniform}          & \text{600 to 1800K}   & \text{planetary atmospheric} & \text{uniform} & \text{600 to 1800K} \\
    \text{temperature (T\textsubscript{p})}                      &                         &                    & \text{temperature (T\textsubscript{p})}        &                & \\ \hline
    \text{stellar temperature}& \text{uniform} & \text{ T\textsubscript{eff}-240 to T\textsubscript{eff}+240K}& \text{stellar temperature }& \text{gaussian}         & $\mu$\text{=T\textsubscript{eff}}, $\sigma$\text{=150K} \\ 
    \text{(T\textsubscript{occ})}   &      &                & \text{(T\textsubscript{star})}   &  & \\ \hline
    \text{stellar heterogeneities}              & \text{uniform}          & \text{T\textsubscript{eff}-3000 to T\textsubscript{eff}+3000K}& \text{stellar heterogeneities} & \text{uniform} & \text{T\textsubscript{eff}-3000 to T\textsubscript{eff}+3000K} \\
    \text{temperature (T\textsubscript{het})}   &                         &  & \text{temperature (T\textsubscript{spot})} &    &   \\\hline
    \text{heterogeneity covering}               & \text{uniform}          & \text{0.0 to 0.5 (\it{WASP-25})}          & \text{heterogeneity covering}                                      & \text{uniform} & \text{0.0 to 0.5 (\it{WASP-25})}  \\ 
    \text{fraction (f\textsubscript{het})}      &                         &  \text{0.0 to 0.068 (\it{WASP-124})}         & \text{fraction (f\textsubscript{spot}) }                           &               &  \text{0.0 to 0.068 (\it{WASP-124})} \\ \hline
    \text{haze amplitude ($a$)}                 & \text{log-uniform}         & \text{-30 to 30}           & \text{scattering factor}                                           & \text{log-uniform} & \text{-10 to 10}\\ \hline
    \text{haze power law (}$\gamma$\text{)} 
    & \text{uniform }         & \text{-14 to 4}            & \text{scattering slope (}$\alpha$\text{)} 
    & \text{uniform} & \text{-4 to 14}\\ \hline
    \text{log cloud absorbing} & \text{uniform} & \text{-80 to 80}        & \text{metallicity (Z/Z}$_{\odot}$)     & \text{log-uniform} & \text{-1 to 3}\\
    \text{cross-section (}$\sigma$\text{\textsubscript{cloud})} &                   &      &                                            &  & \\ \hline
    \text{trace molecules'}       & \text{log-uniform}      & \text{-30 to 0 }             & \text{C/O}                                                         & \text{uniform} & \text{0.05 to 2}\\
    \text{mixing ratios}       &     &            &                                                        &  & \\ \hline
    \text{reference radius factor} ($f$)   & \text{uniform }         & \text{0.8 to 1.2}            & \text{1\,bar, reference radius (R\textsubscript{0})}                       & \text{uniform} & \text{ R\textsubscript{p}-.2R\textsubscript{p} to
    R\textsubscript{p}+.2R\textsubscript{p}}\\ \hline
    \enddata
\tablenotetext{}{\textbf{Note.} These priors were set to allow for a wide parameter space to be surveyed, but contained within physical regimes. Not all parameters were included in each model fit (see Tab.~\ref{tab:W25_W124_LnZs}). We used 5000 live points for all runs. T\textsubscript{eff} is the effective temperature of the host star, which is 5697K and 6258K for WASP-25 and WASP-124, respectively. $\gamma$ is the exponent of the scattering slope power law, where $-4$ is a Rayleigh scattering slope. $\alpha$ is the wavelength dependence of scattering, with 4 being Rayleigh. $f$ is a factor multiplied by the inputted planetary radius to produce the reference radius, i.e., R$_0=f$R$_p$, R$_p$ is the radius of the planet, corresponding to 1.232R$_J$ and 1.337R$_J$ for WASP-25b and WASP-124b, respectively. All stellar and planetary parameters were obtained from Table 1 of \cite{McGruder:2023}.}
\end{deluxetable*}

\begin{acknowledgments}
\edit3{We thank the anonymous referee for helpful comments that improved the manuscript.}
This work has been supported by the National Aeronautics and Space Administration's Exoplanet Research Program via grant No. 20-XRP20\_2.0091.
AJ acknowledges support from ANID -- Millennium  Science  Initiative -- ICN12\_009 and from FONDECYT project 1210718.
JK acknowledges financial support from Imperial College London through an Imperial College Research Fellowship grant.
KNOC acknowledges support from a Ford Foundation Predoctoral Fellowship.
NHA acknowledges support by the National Science Foundation Graduate Research Fellowship under Grant No. DGE1746891.
B.V.R thanks the Heising-Simons Foundation for support.
\end{acknowledgments}

%

\vspace{5mm}
\facilities{Magellan:Baade (IMACS), Smithsonian Institution High Performance Cluster (SI/HPC), and the New Technology Telescope (EFOSC2)}


\software{Astropy \citep{astropy2013}, corner \citep{corner2016}, Matplotlib \citep{matplotlib2007}, NumPy \citep{numpy2020}, Multinest \citep{multinest2009}, PyMultiNest \citep{2014BuchnerPyMultiNest}, SciPy \citep{scipy2020}, batman \citep{Kreidberg2015_batman}, george \citep{Mackey2014_george}} dynesty \citep{Speagle_2020Dynesty}, \texttt{PLATON} \citep{Zhang_2019PLATON}




\bibliographystyle{yahapj}
\bibliography{main}

\begin{thebibliography}{}
\providecommand\natexlab[1]{#1}
\providecommand\JournalTitle[1]{#1}

\bibitem[{{Ahrer} {et~al.}(2023{\natexlab{a}}){Ahrer}, {Wheatley}, {Gandhi},
  {Kirk}, {King}, {Louden}, \& {Welbanks}}]{Ahrer:2023b}
{Ahrer}, E., {Wheatley}, P.~J., {Gandhi}, S., {et~al.} 2023{\natexlab{a}},
  \JournalTitle{arXiv e-prints}, arXiv:2303.07381

\bibitem[{{Ahrer} {et~al.}(2022){Ahrer}, {Wheatley}, {Kirk}, {Gandhi}, {King},
  \& {Louden}}]{Ahrer:2022}
{Ahrer}, E., {Wheatley}, P.~J., {Kirk}, J., {et~al.} 2022,
  \href{http://dx.doi.org/10.1093/mnras/stab3805}{\JournalTitle{\mnras}, 510,
  4857}

\bibitem[{{Ahrer} {et~al.}(2023{\natexlab{b}}){Ahrer}, {Stevenson},
  {Mansfield}, {Moran}, {Brande}, {Morello}, {Murray}, {Nikolov}, {Petit dit de
  la Roche}, {Schlawin}, {Wheatley}, {Zieba}, {Batalha}, {Damiano}, {Goyal},
  {Lendl}, {Lothringer}, {Mukherjee}, {Ohno}, {Batalha}, {Battley}, {Bean},
  {Beatty}, {Benneke}, {Berta-Thompson}, {Carter}, {Cubillos}, {Daylan},
  {Espinoza}, {Gao}, {Gibson}, {Gill}, {Harrington}, {Hu}, {Kreidberg},
  {Lewis}, {Line}, {L{\'o}pez-Morales}, {Parmentier}, {Powell}, {Sing}, {Tsai},
  {Wakeford}, {Welbanks}, {Alam}, {Alderson}, {Allen}, {Anderson}, {Barstow},
  {Bayliss}, {Bell}, {Blecic}, {Bryant}, {Burleigh}, {Carone}, {Casewell},
  {Changeat}, {Chubb}, {Crossfield}, {Crouzet}, {Decin}, {D{\'e}sert},
  {Feinstein}, {Flagg}, {Fortney}, {Gizis}, {Heng}, {Iro}, {Kempton},
  {Kendrew}, {Kirk}, {Knutson}, {Komacek}, {Lagage}, {Leconte},
  {Lustig-Yaeger}, {MacDonald}, {Mancini}, {May}, {Mayne}, {Miguel},
  {Mikal-Evans}, {Molaverdikhani}, {Palle}, {Piaulet}, {Rackham}, {Redfield},
  {Rogers}, {Roy}, {Rustamkulov}, {Shkolnik}, {Sotzen}, {Taylor}, {Tremblin},
  {Tucker}, {Turner}, {de Val-Borro}, {Venot}, \& {Zhang}}]{Ahrer2023}
{Ahrer}, E.-M., {Stevenson}, K.~B., {Mansfield}, M., {et~al.}
  2023{\natexlab{b}},
  \href{http://dx.doi.org/10.1038/s41586-022-05590-4}{\JournalTitle{\nat}, 614,
  653–658}

\bibitem[{{Alam} {et~al.}(2018){Alam}, {Nikolov}, {L{\'o}pez-Morales}, {Sing},
  {Goyal}, {Henry}, {Sanz-Forcada}, {Williamson}, {Evans}, {Wakeford}, {Bruno},
  {Ballester}, {Stevenson}, {Lewis}, {Barstow}, {Bourrier}, {Buchhave},
  {Ehrenreich}, \& {Garc{\'\i}a Mu{\~n}oz}}]{Alam2018}
{Alam}, M.~K., {Nikolov}, N., {L{\'o}pez-Morales}, M., {et~al.} 2018,
  \href{http://dx.doi.org/10.3847/1538-3881/aaee89}{\JournalTitle{\aj}, 156,
  298}

\bibitem[{{Alam} {et~al.}(2020){Alam}, {L{\'o}pez-Morales}, {Nikolov}, {Sing},
  {Henry}, {Baxter}, {D{\'e}sert}, {Barstow}, {Mikal-Evans}, {Bourrier},
  {Lavvas}, {Wakeford}, {Williamson}, {Sanz-Forcada}, {Buchhave}, {Cohen}, \&
  {Garc{\'\i}a Mu{\~n}oz}}]{Alam2020}
{Alam}, M.~K., {L{\'o}pez-Morales}, M., {Nikolov}, N., {et~al.} 2020,
  \href{http://dx.doi.org/10.3847/1538-3881/ab96cb}{\JournalTitle{\aj}, 160,
  51}

\bibitem[{{Alam} {et~al.}(2021){Alam}, {L{\'o}pez-Morales}, {MacDonald},
  {Nikolov}, {Kirk}, {Goyal}, {Sing}, {Wakeford}, {Rathcke}, {Deming},
  {Sanz-Forcada}, {Lewis}, {Barstow}, {Mikal-Evans}, \& {Buchhave}}]{Alam:2021}
{Alam}, M.~K., {L{\'o}pez-Morales}, M., {MacDonald}, R.~J., {et~al.} 2021,
  \href{http://dx.doi.org/10.3847/2041-8213/abd18e}{\JournalTitle{\apjl}, 906,
  L10}

\bibitem[{{Alderson} {et~al.}(2020){Alderson}, {Kirk}, {L{\'o}pez-Morales},
  {Wheatley}, {Skillen}, {Henry}, {McGruder}, {Brogi}, {Louden}, \&
  {King}}]{Alderson:2020}
{Alderson}, L., {Kirk}, J., {L{\'o}pez-Morales}, M., {et~al.} 2020,
  \href{http://dx.doi.org/10.1093/mnras/staa2315}{\JournalTitle{\mnras}, 497,
  5182}

\bibitem[{{Alderson} {et~al.}(2022){Alderson}, {Wakeford}, {MacDonald},
  {Lewis}, {May}, {Grant}, {Sing}, {Stevenson}, {Fowler}, {Goyal}, {Batalha},
  \& {Kataria}}]{Alderson2022}
{Alderson}, L., {Wakeford}, H.~R., {MacDonald}, R.~J., {et~al.} 2022,
  \href{http://dx.doi.org/10.1093/mnras/stac661}{\JournalTitle{\mnras}, 512,
  4185}

\bibitem[{{Alderson} {et~al.}(2023){Alderson}, {Wakeford}, {Alam}, {Batalha},
  {Lothringer}, {Adams Redai}, {Barat}, {Brande}, {Damiano}, {Daylan},
  {Espinoza}, {Flagg}, {Goyal}, {Grant}, {Hu}, {Inglis}, {Lee}, {Mikal-Evans},
  {Ramos-Rosado}, {Roy}, {Wallack}, {Batalha}, {Bean}, {Benneke},
  {Berta-Thompson}, {Carter}, {Changeat}, {Col{\'o}n}, {Crossfield},
  {D{\'e}sert}, {Foreman-Mackey}, {Gibson}, {Kreidberg}, {Line},
  {L{\'o}pez-Morales}, {Molaverdikhani}, {Moran}, {Morello}, {Moses},
  {Mukherjee}, {Schlawin}, {Sing}, {Stevenson}, {Taylor}, {Aggarwal}, {Ahrer},
  {Allen}, {Barstow}, {Bell}, {Blecic}, {Casewell}, {Chubb}, {Crouzet},
  {Cubillos}, {Decin}, {Feinstein}, {Fortney}, {Harrington}, {Heng}, {Iro},
  {Kempton}, {Kirk}, {Knutson}, {Krick}, {Leconte}, {Lendl}, {MacDonald},
  {Mancini}, {Mansfield}, {May}, {Mayne}, {Miguel}, {Nikolov}, {Ohno}, {Palle},
  {Parmentier}, {Petit dit de la Roche}, {Piaulet}, {Powell}, {Rackham},
  {Redfield}, {Rogers}, {Rustamkulov}, {Tan}, {Tremblin}, {Tsai}, {Turner}, {de
  Val-Borro}, {Venot}, {Welbanks}, {Wheatley}, \& {Zhang}}]{Alderson2023}
{Alderson}, L., {Wakeford}, H.~R., {Alam}, M.~K., {et~al.} 2023,
  \href{http://dx.doi.org/10.1038/s41586-022-05591-3}{\JournalTitle{\nat}, 614,
  664–669}

\bibitem[{{Allen} {et~al.}(2022){Allen}, {Espinoza}, {Jord{\'a}n},
  {L{\'o}pez-Morales}, {Apai}, {Rackham}, {Kirk}, {Osip}, {Weaver}, {McGruder},
  {Ceballos}, {Reggiani}, {Brahm}, {Rodler}, {Lewis}, \& {Fraine}}]{Allen2022}
{Allen}, N.~H., {Espinoza}, N., {Jord{\'a}n}, A., {et~al.} 2022,
  \href{http://dx.doi.org/10.3847/1538-3881/ac8b74}{\JournalTitle{\aj}, 164,
  153}

\bibitem[{{Ambikasaran} {et~al.}(2015){Ambikasaran}, {Foreman-Mackey},
  {Greengard}, {Hogg}, \& {O'Neil}}]{Mackey2014_george}
{Ambikasaran}, S., {Foreman-Mackey}, D., {Greengard}, L., {Hogg}, D.~W., \&
  {O'Neil}, M. 2015,
  \href{http://dx.doi.org/10.1109/TPAMI.2015.2448083}{\JournalTitle{IEEE
  Transactions on Pattern Analysis and Machine Intelligence}, 38, 252}

\bibitem[{{Astropy Collaboration} {et~al.}(2013){Astropy Collaboration},
  {Robitaille}, {Tollerud}, {Greenfield}, {Droettboom}, {Bray}, {Aldcroft},
  {Davis}, {Ginsburg}, {Price-Whelan}, {Kerzendorf}, {Conley}, {Crighton},
  {Barbary}, {Muna}, {Ferguson}, {Grollier}, {Parikh}, {Nair}, {Unther},
  {Deil}, {Woillez}, {Conseil}, {Kramer}, {Turner}, {Singer}, {Fox}, {Weaver},
  {Zabalza}, {Edwards}, {Azalee Bostroem}, {Burke}, {Casey}, {Crawford},
  {Dencheva}, {Ely}, {Jenness}, {Labrie}, {Lim}, {Pierfederici}, {Pontzen},
  {Ptak}, {Refsdal}, {Servillat}, \& {Streicher}}]{astropy2013}
{Astropy Collaboration}, {Robitaille}, T.~P., {Tollerud}, E.~J., {et~al.} 2013,
  \href{http://dx.doi.org/10.1051/0004-6361/201322068}{\JournalTitle{\aap},
  558, A33}

\bibitem[{{Barstow} {et~al.}(2020){Barstow}, {Changeat}, {Garland}, {Line},
  {Rocchetto}, \& {Waldmann}}]{Barstow:2020}
{Barstow}, J.~K., {Changeat}, Q., {Garland}, R., {et~al.} 2020,
  \href{http://dx.doi.org/10.1093/mnras/staa548}{\JournalTitle{\mnras}, 493,
  4884}

\bibitem[{{Benneke} \& {Seager}(2013)}]{2013Benneke}
{Benneke}, B., \& {Seager}, S. 2013,
  \href{http://dx.doi.org/10.1088/0004-637X/778/2/153}{\JournalTitle{\apj},
  778, 153}

\bibitem[{{Bixel} {et~al.}(2019){Bixel}, {Rackham}, {Apai}, {Espinoza},
  {L{\'o}pez-Morales}, {Osip}, {Jord{\'a}n}, {McGruder}, \&
  {Weaver}}]{Bixel:2019}
{Bixel}, A., {Rackham}, B.~V., {Apai}, D., {et~al.} 2019,
  \href{http://dx.doi.org/10.3847/1538-3881/aaf9a3}{\JournalTitle{\aj}, 157,
  68}

\bibitem[{{Brown} {et~al.}(2012){Brown}, {Cameron}, {Anderson}, {Enoch},
  {Hellier}, {Maxted}, {Miller}, {Pollacco}, {Queloz}, {Simpson}, {Smalley},
  {Triaud}, {Boisse}, {Bouchy}, {Gillon}, \& {H{\'e}brard}}]{Brown2012}
{Brown}, D.~J.~A., {Cameron}, A.~C., {Anderson}, D.~R., {et~al.} 2012,
  \href{http://dx.doi.org/10.1111/j.1365-2966.2012.20973.x}{\JournalTitle{\mnras},
  423, 1503}

\bibitem[{{Buchner} {et~al.}(2014){Buchner}, {Georgakakis}, {Nandra}, {Hsu},
  {Rangel}, {Brightman}, {Merloni}, {Salvato}, {Donley}, \&
  {Kocevski}}]{2014BuchnerPyMultiNest}
{Buchner}, J., {Georgakakis}, A., {Nandra}, K., {et~al.} 2014,
  \href{http://dx.doi.org/10.1051/0004-6361/201322971}{\JournalTitle{\aap},
  564, A125}

\bibitem[{{Buzzoni} {et~al.}(1984){Buzzoni}, {Delabre}, {Dekker}, {Dodorico},
  {Enard}, {Focardi}, {Gustafsson}, {Nees}, {Paureau}, \&
  {Reiss}}]{Buzzoni:1984}
{Buzzoni}, B., {Delabre}, B., {Dekker}, H., {et~al.} 1984, \JournalTitle{The
  Messenger}, 38, 9

\bibitem[{{Carter} {et~al.}(2020){Carter}, {Nikolov}, {Sing}, {Alam}, {Goyal},
  {Mikal-Evans}, {Wakeford}, {Henry}, {Morrell}, {L{\'o}pez-Morales},
  {Smalley}, {Lavvas}, {Barstow}, {Garc{\'\i}a Mu{\~n}oz}, {Gibson}, \&
  {Wilson}}]{Carter2020}
{Carter}, A.~L., {Nikolov}, N., {Sing}, D.~K., {et~al.} 2020,
  \href{http://dx.doi.org/10.1093/mnras/staa1078}{\JournalTitle{\mnras}, 494,
  5449}

\bibitem[{{Chachan} {et~al.}(2019){Chachan}, {Knutson}, {Gao}, {Kataria},
  {Wong}, {Henry}, {Benneke}, {Zhang}, {Barstow}, {Bean}, {Mikal-Evans},
  {Lewis}, {Mansfield}, {L{\'o}pez-Morales}, {Nikolov}, {Sing}, \&
  {Wakeford}}]{Chachan2019}
{Chachan}, Y., {Knutson}, H.~A., {Gao}, P., {et~al.} 2019,
  \href{http://dx.doi.org/10.3847/1538-3881/ab4e9a}{\JournalTitle{\aj}, 158,
  244}

\bibitem[{{Charbonneau} {et~al.}(2002){Charbonneau}, {Brown}, {Noyes}, \&
  {Gilliland}}]{Charbonneau:2002}
{Charbonneau}, D., {Brown}, T.~M., {Noyes}, R.~W., \& {Gilliland}, R.~L. 2002,
  \href{http://dx.doi.org/10.1086/338770}{\JournalTitle{\apj}, 568, 377}

\bibitem[{{Diamond-Lowe} {et~al.}(2018){Diamond-Lowe}, {Berta-Thompson},
  {Charbonneau}, \& {Kempton}}]{Diamond-Lowe2018}
{Diamond-Lowe}, H., {Berta-Thompson}, Z., {Charbonneau}, D., \& {Kempton}, E.
  M.~R. 2018,
  \href{http://dx.doi.org/10.3847/1538-3881/aac6dd}{\JournalTitle{\aj}, 156,
  42}

\bibitem[{{Dressler} {et~al.}(2011){Dressler}, {Bigelow}, {Hare}, {Sutin},
  {Thompson}, {Burley}, {Epps}, {Oemler}, {Bagish}, {Birk}, {Clardy},
  {Gunnels}, {Kelson}, {Shectman}, \& {Osip}}]{2011Dressler}
{Dressler}, A., {Bigelow}, B., {Hare}, T., {et~al.} 2011,
  \href{http://dx.doi.org/10.1086/658908}{\JournalTitle{\pasp}, 123, 288}

\bibitem[{{Dymont} {et~al.}(2022){Dymont}, {Yu}, {Ohno}, {Zhang}, {Fortney},
  {Thorngren}, \& {Dickinson}}]{Dymont2021}
{Dymont}, A.~H., {Yu}, X., {Ohno}, K., {et~al.} 2022,
  \href{http://dx.doi.org/10.3847/1538-4357/ac7f40}{\JournalTitle{\apj}, 937,
  90}

\bibitem[{{Enoch} {et~al.}(2011){Enoch}, {Cameron}, {Anderson}, {Lister},
  {Hellier}, {Maxted}, {Queloz}, {Smalley}, {Triaud}, {West}, {Brown},
  {Gillon}, {Hebb}, {Lendl}, {Parley}, {Pepe}, {Pollacco}, {Segransan},
  {Simpson}, {Street}, \& {Udry}}]{Enoch2011}
{Enoch}, B., {Cameron}, A.~C., {Anderson}, D.~R., {et~al.} 2011,
  \href{http://dx.doi.org/10.1111/j.1365-2966.2010.17550.x}{\JournalTitle{\mnras},
  410, 1631}

\bibitem[{{Espinoza}(2017)}]{Espinoza:2017}
{Espinoza}, N. 2017, PhD thesis, Pontificia Universidad Cat\'olica de Chile

\bibitem[{{Espinoza} {et~al.}(2019){Espinoza}, {Rackham}, {Jord{\'a}n}, {Apai},
  {L{\'o}pez-Morales}, {Osip}, {Grimm}, {Hoeijmakers}, {Wilson}, {Bixel},
  {McGruder}, {Rodler}, {Weaver}, {Lewis}, {Fortney}, \&
  {Fraine}}]{Espinoza2019}
{Espinoza}, N., {Rackham}, B.~V., {Jord{\'a}n}, A., {et~al.} 2019,
  \href{http://dx.doi.org/10.1093/mnras/sty2691}{\JournalTitle{\mnras}, 482,
  2065}

\bibitem[{{Estrela} {et~al.}(2022){Estrela}, {Swain}, \&
  {Roudier}}]{Estrela:2022}
{Estrela}, R., {Swain}, M.~R., \& {Roudier}, G.~M. 2022,
  \href{http://dx.doi.org/10.3847/2041-8213/aca2aa}{\JournalTitle{\apjl}, 941,
  L5}

\bibitem[{{Estrela} {et~al.}(2021){Estrela}, {Swain}, {Roudier}, {West},
  {Sedaghati}, \& {Valio}}]{Estrela2021}
{Estrela}, R., {Swain}, M.~R., {Roudier}, G.~M., {et~al.} 2021,
  \href{http://dx.doi.org/10.3847/1538-3881/ac0c7c}{\JournalTitle{\aj}, 162,
  91}

\bibitem[{{Feinstein} {et~al.}(2023){Feinstein}, {Radica}, {Welbanks},
  {Murray}, {Ohno}, {Coulombe}, {Espinoza}, {Bean}, {Teske}, {Benneke}, {Line},
  {Rustamkulov}, {Saba}, {Tsiaras}, {Barstow}, {Fortney}, {Gao}, {Knutson},
  {MacDonald}, {Mikal-Evans}, {Rackham}, {Taylor}, {Parmentier}, {Batalha},
  {Berta-Thompson}, {Carter}, {Changeat}, {Dos Santos}, {Gibson}, {Goyal},
  {Kreidberg}, {L{\'o}pez-Morales}, {Lothringer}, {Miguel}, {Molaverdikhani},
  {Moran}, {Morello}, {Mukherjee}, {Sing}, {Stevenson}, {Wakeford}, {Ahrer},
  {Alam}, {Alderson}, {Allen}, {Batalha}, {Bell}, {Blecic}, {Brande},
  {Caceres}, {Casewell}, {Chubb}, {Crossfield}, {Crouzet}, {Cubillos}, {Decin},
  {D{\'e}sert}, {Harrington}, {Heng}, {Henning}, {Iro}, {Kempton}, {Kendrew},
  {Kirk}, {Krick}, {Lagage}, {Lendl}, {Mancini}, {Mansfield}, {May}, {Mayne},
  {Nikolov}, {Palle}, {Petit dit de la Roche}, {Piaulet}, {Powell}, {Redfield},
  {Rogers}, {Roman}, {Roy}, {Nixon}, {Schlawin}, {Tan}, {Tremblin}, {Turner},
  {Venot}, {Waalkes}, {Wheatley}, \& {Zhang}}]{Feinstein2023}
{Feinstein}, A.~D., {Radica}, M., {Welbanks}, L., {et~al.} 2023,
  \href{http://dx.doi.org/10.1038/s41586-022-05674-1}{\JournalTitle{\nat}, 614,
  670–675}

\bibitem[{{Feroz} {et~al.}(2009){Feroz}, {Hobson}, \&
  {Bridges}}]{multinest2009}
{Feroz}, F., {Hobson}, M.~P., \& {Bridges}, M. 2009,
  \href{http://dx.doi.org/10.1111/j.1365-2966.2009.14548.x}{\JournalTitle{\mnras},
  398, 1601}

\bibitem[{{Fisher} \& {Heng}(2018)}]{Fisher:2018}
{Fisher}, C., \& {Heng}, K. 2018,
  \href{http://dx.doi.org/10.1093/mnras/sty2550}{\JournalTitle{\mnras}, 481,
  4698}

\bibitem[{{Fleury} {et~al.}(2019){Fleury}, {Gudipati}, {Henderson}, \&
  {Swain}}]{Fleury:2019}
{Fleury}, B., {Gudipati}, M.~S., {Henderson}, B.~L., \& {Swain}, M. 2019,
  \href{http://dx.doi.org/10.3847/1538-4357/aaf79f}{\JournalTitle{\apj}, 871,
  158}

\bibitem[{Foreman-Mackey(2016)}]{corner2016}
Foreman-Mackey, D. 2016,
  \href{http://dx.doi.org/10.21105/joss.00024}{\JournalTitle{The Journal of
  Open Source Software}, 1, 24}

\bibitem[{{Fu} {et~al.}(2017){Fu}, {Deming}, {Knutson}, {Madhusudhan},
  {Mandell}, \& {Fraine}}]{Fu:2017}
{Fu}, G., {Deming}, D., {Knutson}, H., {et~al.} 2017,
  \href{http://dx.doi.org/10.3847/2041-8213/aa8e40}{\JournalTitle{\apjl}, 847,
  L22}

\bibitem[{{Gao} {et~al.}(2020){Gao}, {Thorngren}, {Lee}, {Fortney}, {Morley},
  {Wakeford}, {Powell}, {Stevenson}, \& {Zhang}}]{Gao:2020}
{Gao}, P., {Thorngren}, D.~P., {Lee}, E. K.~H., {et~al.} 2020,
  \href{http://dx.doi.org/10.1038/s41550-020-1114-3}{\JournalTitle{Nature
  Astronomy}, 4, 951}

\bibitem[{Gibson(2014)}]{Gibson2014_modelAvg}
Gibson, N.~P. 2014,
  \href{http://dx.doi.org/10.1093/mnras/stu1975}{\JournalTitle{Monthly Notices
  of the Royal Astronomical Society}, 445, 3401}

\bibitem[{Harris {et~al.}(2020)Harris, Millman, van~der Walt, Gommers,
  Virtanen, Cournapeau, Wieser, Taylor, Berg, Smith, Kern, Picus, Hoyer, van
  Kerkwijk, Brett, Haldane, del R{\'{i}}o, Wiebe, Peterson,
  G{\'{e}}rard-Marchant, Sheppard, Reddy, Weckesser, Abbasi, Gohlke, \&
  Oliphant}]{numpy2020}
Harris, C.~R., Millman, K.~J., van~der Walt, S.~J., {et~al.} 2020,
  \href{http://dx.doi.org/10.1038/s41586-020-2649-2}{\JournalTitle{Nature},
  585, 357}

\bibitem[{Helling(2019)}]{Helling:2019}
Helling, C. 2019,
  \href{http://dx.doi.org/10.1146/annurev-earth-053018-060401}{\JournalTitle{Annual
  Review of Earth and Planetary Sciences}, 47, 583–606}

\bibitem[{{Heng}(2016)}]{Heng:2016}
{Heng}, K. 2016,
  \href{http://dx.doi.org/10.3847/2041-8205/826/1/L16}{\JournalTitle{\apjl},
  826, L16}

\bibitem[{{Hoffman} {et~al.}(2022){Hoffman}, {Quintana}, {Dotson}, {Col{\'o}n},
  {Barclay}, {Supsinskas}, {Karburn}, {Apai}, {Hedges}, {Rackham}, {Rowe},
  {Christiansen}, {Greene}, {Mason}, {Mosby}, {Espinoza}, {Gilbert}, {Kostov},
  {Lewis}, {Morris}, {Mullally}, {Newton}, {Schlieder}, {Youngblood}, {Foote},
  {Mansfield}, {Stevenson}, {Villanueva}, \& {Pepper}}]{Hoffman:2022}
{Hoffman}, K., {Quintana}, E.~V., {Dotson}, J.~L., {et~al.} 2022,
  \href{http://dx.doi.org/10.1117/12.2629546}{in Society of Photo-Optical
  Instrumentation Engineers (SPIE) Conference Series, Vol. 12180, Space
  Telescopes and Instrumentation 2022: Optical, Infrared, and Millimeter Wave,
  ed. L.~E. {Coyle}, S.~{Matsuura}, \& M.~D. {Perrin}}, 121800C

\bibitem[{Hunter(2007)}]{matplotlib2007}
Hunter, J.~D. 2007,
  \href{http://dx.doi.org/10.1109/MCSE.2007.55}{\JournalTitle{Computing In
  Science \& Engineering}, 9, 90}

\bibitem[{{Jord{\'a}n} {et~al.}(2013){Jord{\'a}n}, {Espinoza}, {Rabus},
  {Eyheramendy}, {Sing}, {D{\'e}sert}, {Bakos}, {Fortney}, {L{\'o}pez-Morales},
  {Maxted}, {Triaud}, \& {Szentgyorgyi}}]{Jordan:2013}
{Jord{\'a}n}, A., {Espinoza}, N., {Rabus}, M., {et~al.} 2013,
  \href{http://dx.doi.org/10.1088/0004-637X/778/2/184}{\JournalTitle{\apj},
  778, 184}

\bibitem[{{Kipping}(2013)}]{Kipping:2013}
{Kipping}, D.~M. 2013,
  \href{http://dx.doi.org/10.1093/mnras/stt1435}{\JournalTitle{\mnras}, 435,
  2152}

\bibitem[{{Kirk} {et~al.}(2019){Kirk}, {L{\'o}pez-Morales}, {Wheatley},
  {Weaver}, {Skillen}, {Louden}, {McCormac}, \& {Espinoza}}]{Kirk:2019}
{Kirk}, J., {L{\'o}pez-Morales}, M., {Wheatley}, P.~J., {et~al.} 2019,
  \href{http://dx.doi.org/10.3847/1538-3881/ab397d}{\JournalTitle{\aj}, 158,
  144}

\bibitem[{{Kirk} {et~al.}(2017){Kirk}, {Wheatley}, {Louden}, {Doyle},
  {Skillen}, {McCormac}, {Irwin}, \& {Karjalainen}}]{Kirk:2017}
{Kirk}, J., {Wheatley}, P.~J., {Louden}, T., {et~al.} 2017,
  \href{http://dx.doi.org/10.1093/mnras/stx752}{\JournalTitle{\mnras}, 468,
  3907}

\bibitem[{{Kirk} {et~al.}(2018){Kirk}, {Wheatley}, {Louden}, {Skillen}, {King},
  {McCormac}, \& {Irwin}}]{Kirk:2018}
---. 2018,
  \href{http://dx.doi.org/10.1093/mnras/stx2826}{\JournalTitle{\mnras}, 474,
  876}

\bibitem[{{Kirk} {et~al.}(2021){Kirk}, {Rackham}, {MacDonald},
  {L{\'o}pez-Morales}, {Espinoza}, {Lendl}, {Wilson}, {Osip}, {Wheatley},
  {Skillen}, {Apai}, {Bixel}, {Gibson}, {Jord{\'a}n}, {Lewis}, {Louden},
  {McGruder}, {Nikolov}, {Rodler}, \& {Weaver}}]{Kirk:2021}
{Kirk}, J., {Rackham}, B.~V., {MacDonald}, R.~J., {et~al.} 2021,
  \href{http://dx.doi.org/10.3847/1538-3881/abfcd2}{\JournalTitle{\aj}, 162,
  34}

\bibitem[{{Knutson} {et~al.}(2011){Knutson}, {Madhusudhan}, {Cowan},
  {Christiansen}, {Agol}, {Deming}, {D{\'e}sert}, {Charbonneau}, {Henry},
  {Homeier}, {Langton}, {Laughlin}, \& {Seager}}]{Knutson2011}
{Knutson}, H.~A., {Madhusudhan}, N., {Cowan}, N.~B., {et~al.} 2011,
  \href{http://dx.doi.org/10.1088/0004-637X/735/1/27}{\JournalTitle{\apj}, 735,
  27}

\bibitem[{{Kreidberg}(2015)}]{Kreidberg2015_batman}
{Kreidberg}, L. 2015,
  \href{http://dx.doi.org/10.1086/683602}{\JournalTitle{\pasp}, 127, 1161}

\bibitem[{{Kreidberg}(2018)}]{Kreidberg2018}
---. 2018, \href{http://dx.doi.org/10.1007/978-3-319-55333-7_100}{in Handbook
  of Exoplanets, ed. H.~J. {Deeg} \& J.~A. {Belmonte}} (Springer, Cham), 100

\bibitem[{{Kulow} {et~al.}(2014){Kulow}, {France}, {Linsky}, \&
  {Loyd}}]{Kulow2014}
{Kulow}, J.~R., {France}, K., {Linsky}, J., \& {Loyd}, R.~O.~P. 2014,
  \href{http://dx.doi.org/10.1088/0004-637X/786/2/132}{\JournalTitle{\apj},
  786, 132}

\bibitem[{{Lecavelier Des Etangs} {et~al.}(2008){Lecavelier Des Etangs},
  {Pont}, {Vidal-Madjar}, \& {Sing}}]{Lecavelier2008}
{Lecavelier Des Etangs}, A., {Pont}, F., {Vidal-Madjar}, A., \& {Sing}, D.
  2008,
  \href{http://dx.doi.org/10.1051/0004-6361:200809388}{\JournalTitle{\aap},
  481, L83}

\bibitem[{{Louden} {et~al.}(2017){Louden}, {Wheatley}, {Irwin}, {Kirk}, \&
  {Skillen}}]{Louden:2017}
{Louden}, T., {Wheatley}, P.~J., {Irwin}, P. G.~J., {Kirk}, J., \& {Skillen},
  I. 2017, \href{http://dx.doi.org/10.1093/mnras/stx984}{\JournalTitle{\mnras},
  470, 742}

\bibitem[{{Marsh}(1989)}]{Marsh1989OptExtract}
{Marsh}, T.~R. 1989,
  \href{http://dx.doi.org/10.1086/132570}{\JournalTitle{\pasp}, 101, 1032}

\bibitem[{{Maxted} {et~al.}(2016){Maxted}, {Anderson}, {Collier Cameron},
  {Delrez}, {Gillon}, {Hellier}, {Jehin}, {Lendl}, {Neveu-VanMalle}, {Pepe},
  {Pollacco}, {Queloz}, {S{\'e}gransan}, {Smalley}, {Smith}, {Southworth},
  {Triaud}, {Udry}, {Wagg}, \& {West}}]{Maxted2016}
{Maxted}, P.~F.~L., {Anderson}, D.~R., {Collier Cameron}, A., {et~al.} 2016,
  \href{http://dx.doi.org/10.1051/0004-6361/201628250}{\JournalTitle{\aap},
  591, A55}

\bibitem[{{May} {et~al.}(2020){May}, {Gardner}, {Rauscher}, \&
  {Monnier}}]{May:2020}
{May}, E.~M., {Gardner}, T., {Rauscher}, E., \& {Monnier}, J.~D. 2020,
  \href{http://dx.doi.org/10.3847/1538-3881/ab5361}{\JournalTitle{\aj}, 159, 7}

\bibitem[{{May} {et~al.}(2018){May}, {Zhao}, {Haidar}, {Rauscher}, \&
  {Monnier}}]{May:2018}
{May}, E.~M., {Zhao}, M., {Haidar}, M., {Rauscher}, E., \& {Monnier}, J.~D.
  2018, \href{http://dx.doi.org/10.3847/1538-3881/aad4a8}{\JournalTitle{\aj},
  156, 122}

\bibitem[{{McGruder} {et~al.}(2023){McGruder}, {L{\'o}pez-Morales}, {Brahm}, \&
  {Jord{\'a}n}}]{McGruder:2023}
{McGruder}, C.~D., {L{\'o}pez-Morales}, M., {Brahm}, R., \& {Jord{\'a}n}, A.
  2023, \href{http://dx.doi.org/10.3847/2041-8213/acb154}{\JournalTitle{\apjl},
  944, L56}

\bibitem[{{McGruder} {et~al.}(2020){McGruder}, {L{\'o}pez-Morales}, {Espinoza},
  {Rackham}, {Apai}, {Jord{\'a}n}, {Osip}, {Alam}, {Bixel}, {Fortney}, {Henry},
  {Kirk}, {Lewis}, {Rodler}, \& {Weaver}}]{McGruder:2020}
{McGruder}, C.~D., {L{\'o}pez-Morales}, M., {Espinoza}, N., {et~al.} 2020,
  \href{http://dx.doi.org/10.3847/1538-3881/abb806}{\JournalTitle{\aj}, 160,
  230}

\bibitem[{{McGruder} {et~al.}(2022){McGruder}, {L{\'o}pez-Morales}, {Kirk},
  {Espinoza}, {Rackham}, {Alam}, {Allen}, {Nikolov}, {Weaver}, {Ortiz
  Ceballos}, {Osip}, {Apai}, {Jord{\'a}n}, \& {Fortney}}]{McGruder:2022}
{McGruder}, C.~D., {L{\'o}pez-Morales}, M., {Kirk}, J., {et~al.} 2022,
  \href{http://dx.doi.org/10.3847/1538-3881/ac7f2e}{\JournalTitle{\aj}, 164,
  134}

\bibitem[{{Moses} {et~al.}(2013){Moses}, {Madhusudhan}, {Visscher}, \&
  {Freedman}}]{moses:2013}
{Moses}, J.~I., {Madhusudhan}, N., {Visscher}, C., \& {Freedman}, R.~S. 2013,
  \href{http://dx.doi.org/10.1088/0004-637X/763/1/25}{\JournalTitle{\apj}, 763,
  25}

\bibitem[{{Moses} {et~al.}(2011){Moses}, {Visscher}, {Fortney}, {Showman},
  {Lewis}, {Griffith}, {Klippenstein}, {Shabram}, {Friedson}, {Marley}, \&
  {Freedman}}]{moses:2011}
{Moses}, J.~I., {Visscher}, C., {Fortney}, J.~J., {et~al.} 2011,
  \href{http://dx.doi.org/10.1088/0004-637X/737/1/15}{\JournalTitle{\apj}, 737,
  15}

\bibitem[{{Nikolov} {et~al.}(2016){Nikolov}, {Sing}, {Gibson}, {Fortney},
  {Evans}, {Barstow}, {Kataria}, \& {Wilson}}]{Nikolov2016}
{Nikolov}, N., {Sing}, D.~K., {Gibson}, N.~P., {et~al.} 2016,
  \href{http://dx.doi.org/10.3847/0004-637X/832/2/191}{\JournalTitle{\apj},
  832, 191}

\bibitem[{{Nikolov} {et~al.}(2015){Nikolov}, {Sing}, {Burrows}, {Fortney},
  {Henry}, {Pont}, {Ballester}, {Aigrain}, {Wilson}, {Huitson}, {Gibson},
  {D{\'e}sert}, {Lecavelier Des Etangs}, {Showman}, {Vidal-Madjar}, {Wakeford},
  \& {Zahnle}}]{Nikolov:2015}
{Nikolov}, N., {Sing}, D.~K., {Burrows}, A.~S., {et~al.} 2015,
  \href{http://dx.doi.org/10.1093/mnras/stu2433}{\JournalTitle{\mnras}, 447,
  463}

\bibitem[{{Nikolov} {et~al.}(2018){Nikolov}, {Sing}, {Fortney}, {Goyal},
  {Drummond}, {Evans}, {Gibson}, {De Mooij}, {Rustamkulov}, {Wakeford},
  {Smalley}, {Burgasser}, {Hellier}, {Helling}, {Mayne}, {Madhusudhan},
  {Kataria}, {Baines}, {Carter}, {Ballester}, {Barstow}, {McCleery}, \&
  {Spake}}]{Nikolov:2018}
{Nikolov}, N., {Sing}, D.~K., {Fortney}, J.~J., {et~al.} 2018,
  \href{http://dx.doi.org/10.1038/s41586-018-0101-7}{\JournalTitle{\nat}, 557,
  526}

\bibitem[{{Nikolov} {et~al.}(2021){Nikolov}, {Maciejewski}, {Constantinou},
  {Madhusudhan}, {Fortney}, {Smalley}, {Carter}, {de Mooij}, {Drummond},
  {Gibson}, {Helling}, {Mayne}, {Mikal-Evans}, {Sing}, \&
  {Wilson}}]{Nikolov2021}
{Nikolov}, N., {Maciejewski}, G., {Constantinou}, S., {et~al.} 2021,
  \href{http://dx.doi.org/10.3847/1538-3881/ac01da}{\JournalTitle{\aj}, 162,
  88}

\bibitem[{{Nikolov} {et~al.}(2022){Nikolov}, {Sing}, {Spake}, {Smalley},
  {Goyal}, {Mikal-Evans}, {Wakeford}, {Rustamkulov}, {Deming}, {Fortney},
  {Carter}, {Gibson}, \& {Mayne}}]{Nikolov2022}
{Nikolov}, N.~K., {Sing}, D.~K., {Spake}, J.~J., {et~al.} 2022,
  \href{http://dx.doi.org/10.1093/mnras/stac1530}{\JournalTitle{\mnras}, 515,
  3037}

\bibitem[{{Ohno} \& {Kawashima}(2020)}]{Ohno2020}
{Ohno}, K., \& {Kawashima}, Y. 2020,
  \href{http://dx.doi.org/10.3847/2041-8213/ab93d7}{\JournalTitle{\apjl}, 895,
  L47}

\bibitem[{{Pizzolato} {et~al.}(2003){Pizzolato}, {Maggio}, {Micela},
  {Sciortino}, \& {Ventura}}]{Pizzolato2003}
{Pizzolato}, N., {Maggio}, A., {Micela}, G., {Sciortino}, S., \& {Ventura}, P.
  2003,
  \href{http://dx.doi.org/10.1051/0004-6361:20021560}{\JournalTitle{\aap}, 397,
  147}

\bibitem[{{Pont} {et~al.}(2013){Pont}, {Sing}, {Gibson}, {Aigrain}, {Henry}, \&
  {Husnoo}}]{Pont2013}
{Pont}, F., {Sing}, D.~K., {Gibson}, N.~P., {et~al.} 2013,
  \href{http://dx.doi.org/10.1093/mnras/stt651}{\JournalTitle{\mnras}, 432,
  2917}

\bibitem[{{Quintana} {et~al.}(2021){Quintana}, {Col{\'o}n}, {Mosby},
  {Schlieder}, {Supsinskas}, {Karburn}, {Dotson}, {Greene}, {Hedges}, {Apai},
  {Barclay}, {Christiansen}, {Espinoza}, {Mullally}, {Gilbert}, {Hoffman},
  {Kostov}, {Lewis}, {Foote}, {Mason}, {Youngblood}, {Morris}, {Newton},
  {Pepper}, {Rackham}, {Rowe}, \& {Stevenson}}]{Quintana:2021}
{Quintana}, E.~V., {Col{\'o}n}, K.~D., {Mosby}, G., {et~al.} 2021,
  \href{http://dx.doi.org/10.48550/arXiv.2108.06438}{\JournalTitle{arXiv
  e-prints}, arXiv:2108.06438}

\bibitem[{{Rackham} {et~al.}(2017){Rackham}, {Espinoza}, {Apai},
  {L{\'o}pez-Morales}, {Jord{\'a}n}, {Osip}, {Lewis}, {Rodler}, {Fraine},
  {Morley}, \& {Fortney}}]{Rackham:2017}
{Rackham}, B., {Espinoza}, N., {Apai}, D., {et~al.} 2017,
  \href{http://dx.doi.org/10.3847/1538-4357/aa4f6c}{\JournalTitle{\apj}, 834,
  151}

\bibitem[{{Rackham} {et~al.}(2019){Rackham}, {Apai}, \&
  {Giampapa}}]{Rackham:2019}
{Rackham}, B.~V., {Apai}, D., \& {Giampapa}, M.~S. 2019,
  \href{http://dx.doi.org/10.3847/1538-3881/aaf892}{\JournalTitle{\aj}, 157,
  96}

\bibitem[{{Rathcke} {et~al.}(2021){Rathcke}, {MacDonald}, {Barstow}, {Goyal},
  {Lopez-Morales}, {Mendon{\c{c}}a}, {Sanz-Forcada}, {Henry}, {Sing}, {Alam},
  {Lewis}, {Chubb}, {Taylor}, {Nikolov}, \& {Buchhave}}]{Rathcke2021}
{Rathcke}, A.~D., {MacDonald}, R.~J., {Barstow}, J.~K., {et~al.} 2021,
  \href{http://dx.doi.org/10.3847/1538-3881/ac0e99}{\JournalTitle{\aj}, 162,
  138}

\bibitem[{{Rustamkulov} {et~al.}(2023){Rustamkulov}, {Sing}, {Mukherjee},
  {May}, {Kirk}, {Schlawin}, {Line}, {Piaulet}, {Carter}, {Batalha}, {Goyal},
  {L{\'o}pez-Morales}, {Lothringer}, {MacDonald}, {Moran}, {Stevenson},
  {Wakeford}, {Espinoza}, {Bean}, {Batalha}, {Benneke}, {Berta-Thompson},
  {Crossfield}, {Gao}, {Kreidberg}, {Powell}, {Cubillos}, {Gibson}, {Leconte},
  {Molaverdikhani}, {Nikolov}, {Parmentier}, {Roy}, {Taylor}, {Turner},
  {Wheatley}, {Aggarwal}, {Ahrer}, {Alam}, {Alderson}, {Allen}, {Banerjee},
  {Barat}, {Barrado}, {Barstow}, {Bell}, {Blecic}, {Brande}, {Casewell},
  {Changeat}, {Chubb}, {Crouzet}, {Daylan}, {Decin}, {D{\'e}sert},
  {Mikal-Evans}, {Feinstein}, {Flagg}, {Fortney}, {Harrington}, {Heng}, {Hong},
  {Hu}, {Iro}, {Kataria}, {Kempton}, {Krick}, {Lendl}, {Lillo-Box}, {Louca},
  {Lustig-Yaeger}, {Mancini}, {Mansfield}, {Mayne}, {Miguel}, {Morello},
  {Ohno}, {Palle}, {Petit dit de la Roche}, {Rackham}, {Radica},
  {Ramos-Rosado}, {Redfield}, {Rogers}, {Shkolnik}, {Southworth}, {Teske},
  {Tremblin}, {Tucker}, {Venot}, {Waalkes}, {Welbanks}, {Zhang}, \&
  {Zieba}}]{Rustamkulov2023}
{Rustamkulov}, Z., {Sing}, D.~K., {Mukherjee}, S., {et~al.} 2023,
  \href{http://dx.doi.org/10.1038/s41586-022-05677-y}{\JournalTitle{\nat}, 614,
  659–663}

\bibitem[{{Sing}(2018)}]{Sing:2018}
{Sing}, D.~K. 2018,
  \href{http://dx.doi.org/10.48550/arXiv.1804.07357}{\JournalTitle{arXiv
  e-prints}, arXiv:1804.07357}

\bibitem[{{Sing} {et~al.}(2012){Sing}, {Huitson}, {Lopez-Morales}, {Pont},
  {D{\'e}sert}, {Ehrenreich}, {Wilson}, {Ballester}, {Fortney}, {Lecavelier des
  Etangs}, \& {Vidal-Madjar}}]{Sing2012}
{Sing}, D.~K., {Huitson}, C.~M., {Lopez-Morales}, M., {et~al.} 2012,
  \href{http://dx.doi.org/10.1111/j.1365-2966.2012.21938.x}{\JournalTitle{\mnras},
  426, 1663}

\bibitem[{{Sing} {et~al.}(2016){Sing}, {Fortney}, {Nikolov}, {Wakeford},
  {Kataria}, {Evans}, {Aigrain}, {Ballester}, {Burrows}, {Deming},
  {D{\'e}sert}, {Gibson}, {Henry}, {Huitson}, {Knutson}, {Lecavelier Des
  Etangs}, {Pont}, {Showman}, {Vidal-Madjar}, {Williamson}, \&
  {Wilson}}]{Sing:2016}
{Sing}, D.~K., {Fortney}, J.~J., {Nikolov}, N., {et~al.} 2016,
  \href{http://dx.doi.org/10.1038/nature16068}{\JournalTitle{\nat}, 529, 59}

\bibitem[{{Southworth} {et~al.}(2014){Southworth}, {Hinse}, {Burgdorf}, {Calchi
  Novati}, {Dominik}, {Galianni}, {Gerner}, {Giannini}, {Gu}, {Hundertmark},
  {J{\o}rgensen}, {Juncher}, {Kerins}, {Mancini}, {Rabus}, {Ricci},
  {Sch{\"a}fer}, {Skottfelt}, {Tregloan-Reed}, {Wang}, {Wertz}, {Alsubai},
  {Andersen}, {Bozza}, {Bramich}, {Browne}, {Ciceri}, {D'Ago}, {Damerdji},
  {Diehl}, {Dodds}, {Elyiv}, {Fang}, {Finet}, {Figuera Jaimes}, {Hardis},
  {Harps{\o}e}, {Jessen-Hansen}, {Kains}, {Kjeldsen}, {Korhonen}, {Liebig},
  {Lund}, {Lundkvist}, {Mathiasen}, {Penny}, {Popovas}, {Prof.}, {Rahvar},
  {Sahu}, {Scarpetta}, {Schmidt}, {Sch{\"o}nebeck}, {Snodgrass}, {Street},
  {Surdej}, {Tsapras}, \& {Vilela}}]{Southworth2014}
{Southworth}, J., {Hinse}, T.~C., {Burgdorf}, M., {et~al.} 2014,
  \href{http://dx.doi.org/10.1093/mnras/stu1492}{\JournalTitle{\mnras}, 444,
  776}

\bibitem[{{Speagle}(2020)}]{Speagle_2020Dynesty}
{Speagle}, J.~S. 2020,
  \href{http://dx.doi.org/10.1093/mnras/staa278}{\JournalTitle{\mnras}, 493,
  3132}

\bibitem[{{Spyratos} {et~al.}(2021){Spyratos}, {Nikolov}, {Southworth},
  {Constantinou}, {Madhusudhan}, {Carter}, {de Mooij}, {Fortney}, {Gibson},
  {Goyal}, {Helling}, {Mayne}, \& {Mikal-Evans}}]{Spyratos2021}
{Spyratos}, P., {Nikolov}, N., {Southworth}, J., {et~al.} 2021,
  \href{http://dx.doi.org/10.1093/mnras/stab1847}{\JournalTitle{\mnras}, 506,
  2853}

\bibitem[{{Stevenson}(2016)}]{Stevenson:2016}
{Stevenson}, K.~B. 2016,
  \href{http://dx.doi.org/10.3847/2041-8205/817/2/L16}{\JournalTitle{\apjl},
  817, L16}

\bibitem[{{Tinetti} {et~al.}(2018){Tinetti}, {Drossart}, {Eccleston},
  {Hartogh}, {Heske}, {Leconte}, {Micela}, {Ollivier}, {Pilbratt}, {Puig},
  {Turrini}, {Vandenbussche}, {Wolkenberg}, {Beaulieu}, {Buchave}, {Ferus},
  {Griffin}, {Guedel}, {Justtanont}, {Lagage}, {Machado}, {Malaguti}, {Min},
  {N{\o}rgaard-Nielsen}, {Rataj}, {Ray}, {Ribas}, {Swain}, {Szabo}, {Werner},
  {Barstow}, {Burleigh}, {Cho}, {Coud{\'e} du Foresto}, {Coustenis}, {Decin},
  {Encrenaz}, {Galand}, {Gillon}, {Helled}, {Morales}, {Garc{\'\i}a Mu{\~n}oz},
  {Moneti}, {Pagano}, {Pascale}, {Piccioni}, {Pinfield}, {Sarkar}, {Selsis},
  {Tennyson}, {Triaud}, {Venot}, {Waldmann}, {Waltham}, {Wright}, {Amiaux},
  {Augu{\`e}res}, {Berth{\'e}}, {Bezawada}, {Bishop}, {Bowles}, {Coffey},
  {Colom{\'e}}, {Crook}, {Crouzet}, {Da Peppo}, {Sanz}, {Focardi}, {Frericks},
  {Hunt}, {Kohley}, {Middleton}, {Morgante}, {Ottensamer}, {Pace}, {Pearson},
  {Stamper}, {Symonds}, {Rengel}, {Renotte}, {Ade}, {Affer}, {Alard}, {Allard},
  {Altieri}, {Andr{\'e}}, {Arena}, {Argyriou}, {Aylward}, {Baccani}, {Bakos},
  {Banaszkiewicz}, {Barlow}, {Batista}, {Bellucci}, {Benatti}, {Bernardi},
  {B{\'e}zard}, {Blecka}, {Bolmont}, {Bonfond}, {Bonito}, {Bonomo}, {Brucato},
  {Brun}, {Bryson}, {Bujwan}, {Casewell}, {Charnay}, {Pestellini}, {Chen},
  {Ciaravella}, {Claudi}, {Cl{\'e}dassou}, {Damasso}, {Damiano}, {Danielski},
  {Deroo}, {Di Giorgio}, {Dominik}, {Doublier}, {Doyle}, {Doyon}, {Drummond},
  {Duong}, {Eales}, {Edwards}, {Farina}, {Flaccomio}, {Fletcher}, {Forget},
  {Fossey}, {Fr{\"a}nz}, {Fujii}, {Garc{\'\i}a-Piquer}, {Gear}, {Geoffray},
  {G{\'e}rard}, {Gesa}, {Gomez}, {Graczyk}, {Griffith}, {Grodent}, {Guarcello},
  {Gustin}, {Hamano}, {Hargrave}, {Hello}, {Heng}, {Herrero}, {Hornstrup},
  {Hubert}, {Ida}, {Ikoma}, {Iro}, {Irwin}, {Jarchow}, {Jaubert}, {Jones},
  {Julien}, {Kameda}, {Kerschbaum}, {Kervella}, {Koskinen}, {Krijger}, {Krupp},
  {Lafarga}, {Landini}, {Lellouch}, {Leto}, {Luntzer}, {Rank-L{\"u}ftinger},
  {Maggio}, {Maldonado}, {Maillard}, {Mall}, {Marquette}, {Mathis}, {Maxted},
  {Matsuo}, {Medvedev}, {Miguel}, {Minier}, {Morello}, {Mura}, {Narita},
  {Nascimbeni}, {Nguyen Tong}, {Noce}, {Oliva}, {Palle}, {Palmer}, {Pancrazzi},
  {Papageorgiou}, {Parmentier}, {Perger}, {Petralia}, {Pezzuto},
  {Pierrehumbert}, {Pillitteri}, {Piotto}, {Pisano}, {Prisinzano}, {Radioti},
  {R{\'e}ess}, {Rezac}, {Rocchetto}, {Rosich}, {Sanna}, {Santerne}, {Savini},
  {Scandariato}, {Sicardy}, {Sierra}, {Sindoni}, {Skup}, {Snellen}, {Sobiecki},
  {Soret}, {Sozzetti}, {Stiepen}, {Strugarek}, {Taylor}, {Taylor}, {Terenzi},
  {Tessenyi}, {Tsiaras}, {Tucker}, {Valencia}, {Vasisht}, {Vazan}, {Vilardell},
  {Vinatier}, {Viti}, {Waters}, {Wawer}, {Wawrzaszek}, {Whitworth}, {Yung},
  {Yurchenko}, {Zapatero Osorio}, {Zellem}, {Zingales}, \&
  {Zwart}}]{Tinetti2018}
{Tinetti}, G., {Drossart}, P., {Eccleston}, P., {et~al.} 2018,
  \href{http://dx.doi.org/10.1007/s10686-018-9598-x}{\JournalTitle{Experimental
  Astronomy}, 46, 135}

\bibitem[{{Todorov} {et~al.}(2019){Todorov}, {D{\'e}sert}, {Huitson}, {Bean},
  {Panwar}, {de Matos}, {Stevenson}, {Fortney}, \& {Bergmann}}]{Todorov2019}
{Todorov}, K.~O., {D{\'e}sert}, J.-M., {Huitson}, C.~M., {et~al.} 2019,
  \href{http://dx.doi.org/10.1051/0004-6361/201935364}{\JournalTitle{\aap},
  631, A169}

\bibitem[{{Trotta}(2008)}]{2008Trotta}
{Trotta}, R. 2008,
  \href{http://dx.doi.org/10.1080/00107510802066753}{\JournalTitle{Contemporary
  Physics}, 49, 71}

\bibitem[{{Tsai} {et~al.}(2022){Tsai}, {Lee}, {Powell}, {Gao}, {Zhang},
  {Moses}, {H{\'e}brard}, {Venot}, {Parmentier}, {Jordan}, {Hu}, {Alam},
  {Alderson}, {Batalha}, {Bean}, {Benneke}, {Bierson}, {Brady}, {Carone},
  {Carter}, {Chubb}, {Inglis}, {Leconte}, {Lopez-Morales}, {Miguel},
  {Molaverdikhani}, {Rustamkulov}, {Sing}, {Stevenson}, {Wakeford}, {Yang},
  {Aggarwal}, {Baeyens}, {Barat}, {Borro}, {Daylan}, {Fortney}, {France},
  {Goyal}, {Grant}, {Kirk}, {Kreidberg}, {Louca}, {Moran}, {Mukherjee},
  {Nasedkin}, {Ohno}, {Rackham}, {Redfield}, {Taylor}, {Tremblin}, {Visscher},
  {Wallack}, {Welbanks}, {Youngblood}, {Ahrer}, {Batalha}, {Behr},
  {Berta-Thompson}, {Blecic}, {Casewell}, {Crossfield}, {Crouzet}, {Cubillos},
  {Decin}, {D{\'e}sert}, {Feinstein}, {Gibson}, {Harrington}, {Heng},
  {Henning}, {Kempton}, {Krick}, {Lagage}, {Lendl}, {Line}, {Lothringer},
  {Mansfield}, {Mayne}, {Mikal-Evans}, {Palle}, {Schlawin}, {Shorttle},
  {Wheatley}, \& {Yurchenko}}]{Tsai2022}
{Tsai}, S.-M., {Lee}, E. K.~H., {Powell}, D., {et~al.} 2022,
  \href{http://dx.doi.org/10.48550/arXiv.2211.10490}{\JournalTitle{arXiv
  e-prints}, arXiv:2211.10490}

\bibitem[{{Tsiaras} {et~al.}(2016){Tsiaras}, {Waldmann}, {Rocchetto}, {Varley},
  {Morello}, {Damiano}, \& {Tinetti}}]{Tsiaras2016}
{Tsiaras}, A., {Waldmann}, I.~P., {Rocchetto}, M., {et~al.} 2016,
  \href{http://dx.doi.org/10.3847/0004-637X/832/2/202}{\JournalTitle{\apj},
  832, 202}

\bibitem[{{Tsiaras} {et~al.}(2018){Tsiaras}, {Waldmann}, {Zingales},
  {Rocchetto}, {Morello}, {Damiano}, {Karpouzas}, {Tinetti}, {McKemmish},
  {Tennyson}, \& {Yurchenko}}]{Tsiaras:2018}
{Tsiaras}, A., {Waldmann}, I.~P., {Zingales}, T., {et~al.} 2018,
  \href{http://dx.doi.org/10.3847/1538-3881/aaaf75}{\JournalTitle{\aj}, 155,
  156}

\bibitem[{Virtanen {et~al.}(2020)Virtanen, Gommers, Oliphant, Haberland, Reddy,
  Cournapeau, Burovski, Peterson, Weckesser, Bright, {van der Walt}, Brett,
  Wilson, Millman, Mayorov, Nelson, Jones, Kern, Larson, Carey, Polat, Feng,
  Moore, {VanderPlas}, Laxalde, Perktold, Cimrman, Henriksen, Quintero, Harris,
  Archibald, Ribeiro, Pedregosa, {van Mulbregt}, \& {SciPy 1.0
  Contributors}}]{scipy2020}
Virtanen, P., Gommers, R., Oliphant, T.~E., {et~al.} 2020,
  \href{http://dx.doi.org/10.1038/s41592-019-0686-2}{\JournalTitle{Nature
  Methods}, 17, 261}

\bibitem[{{Wakeford} {et~al.}(2020){Wakeford}, {Sing}, {Stevenson}, {Lewis},
  {Pirzkal}, {Wilson}, {Goyal}, {Kataria}, {Mikal-Evans}, {Nikolov}, \&
  {Spake}}]{Wakeford2020}
{Wakeford}, H.~R., {Sing}, D.~K., {Stevenson}, K.~B., {et~al.} 2020,
  \href{http://dx.doi.org/10.3847/1538-3881/ab7b78}{\JournalTitle{\aj}, 159,
  204}

\bibitem[{{Weaver} {et~al.}(2020){Weaver}, {L{\'o}pez-Morales}, {Espinoza},
  {Rackham}, {Osip}, {Apai}, {Jord{\'a}n}, {Bixel}, {Lewis}, {Alam}, {Kirk},
  {McGruder}, {Rodler}, \& {Fienco}}]{Weaver:2020}
{Weaver}, I.~C., {L{\'o}pez-Morales}, M., {Espinoza}, N., {et~al.} 2020,
  \href{http://dx.doi.org/10.3847/1538-3881/ab55da}{\JournalTitle{\aj}, 159,
  13}

\bibitem[{{Weaver} {et~al.}(2021){Weaver}, {L{\'o}pez-Morales}, {Alam},
  {Espinoza}, {Rackham}, {Goyal}, {MacDonald}, {Lewis}, {Apai}, {Bixel},
  {Jord{\'a}n}, {Kirk}, {McGruder}, \& {Osip}}]{Weaver:2021}
{Weaver}, I.~C., {L{\'o}pez-Morales}, M., {Alam}, M.~K., {et~al.} 2021,
  \href{http://dx.doi.org/10.3847/1538-3881/abf652}{\JournalTitle{\aj}, 161,
  278}

\bibitem[{{Wright} {et~al.}(2011){Wright}, {Drake}, {Mamajek}, \&
  {Henry}}]{Wright2011}
{Wright}, N.~J., {Drake}, J.~J., {Mamajek}, E.~E., \& {Henry}, G.~W. 2011,
  \href{http://dx.doi.org/10.1088/0004-637X/743/1/48}{\JournalTitle{\apj}, 743,
  48}

\bibitem[{{Wright} {et~al.}(2013){Wright}, {Drake}, {Mamajek}, \&
  {Henry}}]{Wright2013}
---. 2013,
  \href{http://dx.doi.org/10.1002/asna.201211764}{\JournalTitle{Astronomische
  Nachrichten}, 334, 151}

\bibitem[{{Yan} {et~al.}(2020){Yan}, {Espinoza}, {Molaverdikhani}, {Henning},
  {Mancini}, {Mallonn}, {Rackham}, {Apai}, {Jord{\'a}n}, {Molli{\`e}re},
  {Chen}, {Carone}, \& {Reiners}}]{Yan2020}
{Yan}, F., {Espinoza}, N., {Molaverdikhani}, K., {et~al.} 2020,
  \href{http://dx.doi.org/10.1051/0004-6361/201937265}{\JournalTitle{\aap},
  642, A98}

\bibitem[{{Zacharias} {et~al.}(2013){Zacharias}, {Finch}, {Girard}, {Henden},
  {Bartlett}, {Monet}, \& {Zacharias}}]{2013UCAC4}
{Zacharias}, N., {Finch}, C.~T., {Girard}, T.~M., {et~al.} 2013,
  \href{http://dx.doi.org/10.1088/0004-6256/145/2/44}{\JournalTitle{\aj}, 145,
  44}

\bibitem[{{Zhang} {et~al.}(2019){Zhang}, {Chachan}, {Kempton}, \&
  {Knutson}}]{Zhang_2019PLATON}
{Zhang}, M., {Chachan}, Y., {Kempton}, E. M.~R., \& {Knutson}, H.~A. 2019,
  \href{http://dx.doi.org/10.1088/1538-3873/aaf5ad}{\JournalTitle{\pasp}, 131,
  034501}

\end{thebibliography}



\end{document}